\documentclass[11pt]{article}
\usepackage{amsmath,amsfonts}
\usepackage{slashed}
\usepackage[dvipdfmx]{graphicx,color}
\usepackage{braket}
\usepackage[nosort]{cite}
\usepackage{ulem}
\usepackage{mathtools}
\usepackage{xcolor}
\usepackage[skins,theorems]{tcolorbox}
\usepackage[makeroom]{cancel}
\usepackage[multiple]{footmisc}

\def\numberbysection{\@addtoreset{equation}{section}
\def\theequation{\arabic{section}.\arabic{equation}}}

\newcommand{\tr}{\, \textrm{tr}}

\newcommand{\E}{{\cal E}}
\newcommand{\cL}{{\cal L}}

\newcommand{\cG}{{\cal G}}
\newcommand{\cO}{{\cal O}}

\newcommand{\be}{\begin{equation}}
\newcommand{\ee}{\end{equation}}
\newcommand{\bea}{\setlength\arraycolsep{2pt} \begin{eqnarray}}
\newcommand{\eea}{\end{eqnarray}}
\newcommand{\ba}{\begin{array}}
\newcommand{\ea}{\end{array}}
\newcommand{\bn}{\begin{align}}
\newcommand{\en}{\end{align}}

\newcommand{\eq}[1]{\eqref{#1}}
\newcommand{\nn}{\nonumber}
\newcommand{\w}[1]{\\[0.#1cm]}

\newcommand{\bpsi}{{\bar\psi}}

\newcommand{\al}{\alpha}
\newcommand{\del}{\delta}
\newcommand{\e}{\epsilon}

\newcommand{\g}{\gamma}

\newcommand{\la}{\lambda}

\newcommand{\m}{\mu}
\newcommand{\n}{\nu}
\newcommand{\om}{\omega}

\newcommand{\pd}{\partial}
\newcommand{\rh}{\rho}
\newcommand{\s}{\sigma}
\newcommand{\ta}{\tau}

\newcommand{\vp}{\varphi}

\usepackage{ifthen}
\reversemarginpar

\numberwithin{equation}{section}

\newcommand{\ve}{\varepsilon}
 
\newcommand{\tG}{{\widetilde G}}
\newcommand{\tcG}{{\widetilde{\cal G}}}
\newcommand{\tH}{{\widetilde H}}
\newcommand{\nonu}{\nonumber \\[.5mm]}

\begin{document}

\allowdisplaybreaks

\thispagestyle{empty}

{\flushright  MI-HET-786 \\
STUPP-22-259 \\[15mm]}

\begin{center}

{ \Large \bf Dualization of Higher Derivative Heterotic\w2 Supergravities in 6D and 10D}\\[5mm]

\vspace{6mm}
\normalsize
{\large  Hao-Yuan Chang${}^{1}$, Ergin Sezgin${}^{1}$ and Yoshiaki Tanii${}^2$}

\vspace{10mm}

${}^4${\it Mitchell Institute for Fundamental Physics and Astronomy\\ Texas A\&M University
College Station, TX 77843, USA}

\vskip 1 em 

${}^2${\it Division of Material Science, Graduate School of Science and Engineering\\ Saitama University, Saitama 338-8570, Japan}

\vspace{10mm}

\hrule

\vspace{5mm}

\begin{tabular}{p{14cm}}

There exist two  four-derivative extensions of $N=(1,0)$ supergravity in six dimensions.
A particular combination of them is known to dualize to the analog of the the Bergshoeff-de Roo (BdR) action in $10D$. Here we first show that the two extensions are not related to each other by any field redefinitions. Next, we dualize them separately thereby obtaining a two parameter dual theory. This is done directly at the level of the action, thus avoiding the laborious method of integrating equations of motion of the dualized theory into an action. To explore whether a similar phenomenon exists in $10D$, we study the dualization of the BdR action in $10D$ in detail. We find an obstacle in the separation of the result into a sum of two independent invariants because of the presence of terms which do not lift from $6D$ to $10D$. We also compare the dual of the BdR action with an existing result obtained in superspace. We find that the bosonic actions agree modulo field redefinitions.

\end{tabular}

\vspace{6mm}
\hrule
\end{center}

\newpage

\setcounter{tocdepth}{2}

\tableofcontents

\medskip

\newpage

%%%%%%%%%%%%%%%%%%%%%%%%%%%%
\section{Introduction}
%%%%%%%%%%%%%%%%%%%%%%%%%%%%

Low energy limit of heterotic string was studied long ago either by computing tree-amplitudes and seeking an action that produces them~\cite{Gross:1986mw,Cai:1986sa,Chang:1986ac}, or from the requirement of superconformal symmetry of the worldsheet sigma model~\cite{Metsaev:1987zx}. In the latter approach, and to first order in $\alpha'$, it was found that the bosonic part of the action agrees with that of Bergshoeff-de Roo (BdR) action ~\cite{Bergshoeff:1989de}, which was obtained from supersymmetry of a four-derivative extension of heterotic supergravity. Given the presence of Lorentz Chern-Simons term in the three-form field strength, which is necessary for the anomaly cancellation, supersymmetry requires order by order deformation of the action and supersymmetry transformations to infinite order in a derivative expansion. As is well known, string theory fixes all the relative coefficients. Nonetheless, since it is still not known if string theory is a unique UV completion of quantum supergravity, it is useful to understand full consequences of supersymmetry, and attempt to determine the coefficients in the derivative expansion from first principles such as locality, causality and high energy unitarity. 

The construction of the BdR action relies on a trick that exploits the similarity in the supertransformation rules of Yang-Mills multiplet and a composite multiplet made of properly supercovariantized Lorentz spin connection and the gravitino curvature~\cite{Bergshoeff:1987rb,Bergshoeff:1988nn}.  It is well known that this approach cannot produce the most general supersymmetric action at the order of eight derivatives and beyond. Nonetheless one may ask whether the BdR action provides a unique four-derivative extension. 
To rigorously prove this, an analysis of the most general superspace constraints, or the Noether procedure in its most general form that does not rely on the `Lorentz from Yang-Mills" trick may be needed. We shall come back to the deformation problem in superspace in Section 5, but let us note that the construction of higher derivative supergravities by Noether procedure is notoriously difficult. For early attempts for direct Noether procedure, see~\cite{Romans:1985xd} where the Gauss-Bonnet combination of curvature squared terms, and~\cite{Han:1985xz} where Riemann-squared terms were considered. Partial results were also obtained in~\cite{Chamseddine:1986gj} where the dual six-form formulation with Gauss-Bonnet terms was considered.  This version of supergravity describes the low energy limit of five-brane, and  provides a lowest order description of string-five brane duality~\cite{Strominger:1990et,Duff:1990ya}. Interestingly enough, a two parameter extension containing Riemann-squared and the Gauss-Bonnet terms was found in~\cite{Bergshoeff:1986wc}, where supersymmetry variations that are independent of the seven-form field strength were considered. Whether these two parameters get related to each other upon taking into account all variations remains to be seen.

Thanks to the availability of superconformal tensor calculus and off-shell formulation~\cite{Bergshoeff:1985mz,Coomans:2011ih}, the study of higher derivative extensions of $N=(1,0), 6D$ supergravity\footnote{What we refer to as  $N=(1,0), 6D$ supergravity actually consists of the irreducible supergravity and a single tensor multiplet. This reducible multiplet has the same field content as $N=(1,0), 10D$ supergravity, and similar couplings to Yang-Mils, and therefore we shall refer to it as heterotic supergravity in $6D$.} are more accessible~\cite{Bergshoeff:2012ax,Ozkan:2013cab,Butter:2018wss}. It is known that the theory admits two distinct four-derivative extensions~\cite{Bergshoeff:2012ax,Butter:2018wss}, and that a particular combination of them arises from the dualization of the BdR action in $6D$~\cite{Liu:2013dna}. This shows that there does exist a four-derivative extension of heterotic supergravity in $6D$ which cannot be obtained by the ``Lorentz from Yang-Mills'' trick~\cite{Bergshoeff:1987rb,Bergshoeff:1988nn}. 
The similarity between heterotic supergravities in $6D$ and the connection afforded by toroidal compactification followed by a consistent truncation, motivates a closer look at the structure of the four-derivative extensions, and the nature of the duality transformations that take into account these extensions. 

The main goal of this paper is to first show that the two four-derivative extensions of heterotic supergravity in $6D$ cannot be related to each other by any field redefinition. One of them has a Riemann$^2$ term~\cite{Bergshoeff:1986wc,Bergshoeff:2012ax}, and the other one has the Gauss-Bonnet combination~\cite{Butter:2018wss} $(R^{\m\n\rh\s} R_{\m\n\rh\s}-4R^{\m\n}R_{\m\n}+R^2)$. The fact that the $R_{\m\n}^2$ and $R^2$ terms can be removed by redefinition of the metric does not imply on-shell equivalence of the two invariants  because several terms involving the three-form field strength, the dilaton and fermions do not coincide.  Next, we shall dualize both extensions separately, thereby obtaining a two parameter $(\alpha,\gamma)$ dual theory with Lagrangian $\cL^{\rm dual}_{\alpha,\gamma}$. We will do so at the level of the action, by adding a suitable Lagrange multiplier term that involves a dual two-form potential $C$, and integrating out the original field strength $H=dB$. We thus avoid the laborious method of integrating equations of motion of the dualized theory into an action, which furthermore may harbor some ambiguities.  
Setting $\gamma=\alpha$ remarkably gives the $6D$ analog of the BdR action. This is in agreement with the dualization performed in the opposite direction by integrating out the dualized equations of motion into an action~\cite{Liu:2013dna}. Setting $\gamma=-\alpha$ instead gives a four-derivative extension in which the curvature-squared terms are $\left(R_{\m\n} R^{\m\n} -\tfrac14 R^2\right)$. These particular terms can be removed by field redefinitions, at the expense of complicating the supersymmetry transformation rules. 

In order to explore whether a similar phenomenon exists in $10D$, we shall examine the dualization of the BdR action in $10D$ in detail. This time the Lagrange multiplier term involves the dual six-form potential $B$, and we integrate over the Lorentz Chern-Simons modified field strength ${\cal G}= dC + \alpha' \omega_L$.  This dualization was performed long ago in ~\cite{Bergshoeff:1990ax,BdR90} but here we provide a closer look at this dualization in the brane frame, working out in detail the dependence on $H=dB$, and explore the possibility of extracting two distinct invariants in analogy with the ${\rm Riem}^2$ and Gauss-Bonnet invariants that exist in $6D$. 
We will show that there is an obstacle in finding such a separation  because of the presence of terms which do not lift from $6D$ to $10D$. We also compare the dual of the BdR action with an existing result obtained in superspace~\cite{Terentev:1993wm,Terentev:1994br,Zyablyuk:1994xk,Saulina:1995eq,Saulina:1996vn}. We find that the bosonic actions agree modulo field redefinitions, but a full comparison in the fermionic sector as well as the supersymmetry transformations remain to be investigated further.

This paper is organized as follows. In Section 2, we recall two four-derivative off-shell invariants of $N=(1,0), 6D$ supergravity, and how to go on-shell. Next, we show that the resulting invariants cannot be related to each other by any field redefinitions. We also highlight the difference between these invariants and the $6D$ analog of BdR  invariant\footnote{In the context of on-shell supergravity, by ``invariant" we  mean a Lagrangian which always contains the two-derivative supergravity plus higher derivative terms with an overall arbitrary constant parameter, and therefore invariant up to a given order in the deformation parameter, such as $\alpha'$, that counts the number of derivatives.}. In Section 3, we dualize the two-parameter Lagrangian $\cL_{\alpha,\gamma}$ and show that for $\gamma=\alpha$ that the result agrees with that of the $6D$ analog of BdR action, upon performing certain field redefinitions. In Section 4, we dualize the BdR action in $10D$ and discuss the obstacle in interpreting the result as a particular combination of two distinct invariants. We shall also compare the dual Lagrangian with that obtained in superspace in \cite{Saulina:1996vn} in Section 5, where we shall also comment on aspects of the superspace formulation in the two-form formulation. Our results are summarized and future directions are noted in the Conclusions, conventions and some identities are collected in Appendix A, and the dimensional reduction of the dual of the BdR action in $10D$ on $T^4$ is described in Appendix B.

%%%%%%%%%%%%%%%%%%%%%%%%%%%%%%%%%%%%%%%%%%%%%%%%%%%%%%%%%%%%%
\section{Higher derivative heterotic supergravity in $6D$ }
%%%%%%%%%%%%%%%%%%%%%%%%%%%%%%%%%%%%%%%%%%%%%%%%%%%%%%%%%%%%%

The known four-derivative $N=(1,0)$ supergravities in $6D$ are as follows. A Riemann-squared invariant was constructed long ago in \cite{Bergshoeff:1986wc}. A Gauss-Bonnet invariant was  constructed partially in \cite{Bergshoeff:1986wc}, and its construction was completed in \cite{Butter:2018wss}. The square of the scalar curvature was given in \cite{Ozkan:2013cab} but its bosonic part can be easily shown be completely removable by a redefinition of the dilaton field. For completeness, we shall also consider the $6D$ analog of the $10D$ Bergshoeff-de Roo (BdR) Lagrangian, which may naively be considered to be the dual formulation of the Riemann-squared invariant mentioned above, but this is not so \cite{Liu:2013dna}, as we shall explain it in detail later. We shall comment on the superspace formulations of all these invariants in Section 5.

\subsection{The two parameter higher derivative Lagrangian $\cL_{\alpha,\gamma}$ }
%%%%%%%%%%%%%%%%%%%%%%%%%%%%%%%%%%%%%%%%%%%%%%%%%%%%%%%%%%%%%%%%%%%%%%%%%%%%%%%%%%

An off-shell $N=(1,0), 6D$ supergravity \cite{Bergshoeff:1985mz} and an off-shell Riemann-squared invariant  \cite{Bergshoeff:1986wc} have been known for sometime. The construction of another off-shell invariant containing the Gauss-Bonnet combination of curvature-squared terms instead was partially achieved in \cite{Bergshoeff:1986wc} and it was completed in \cite{Butter:2018wss}. The field content of the $48+48$ degree of freedom off-shell Poincar\'e multiplet is \cite{Bergshoeff:1985mz}
\be
\left( e_\m{}^a, B_{\m\n},\ \vp,\ V_\m,\ Z_\m,\ E_{\m\n\rh\s},\ \psi_\m^i,\ \chi^i \right)
\ee
where $B,V, E$ are form potentials with associated gauge symmetries, $Z_\m$ is a complex vector field, $\vp$ is the dilaton, and the spinors with $i=1,2$ are symplectic Majorana-Weyl. Adding these three off-shell invariants\footnote{ We shall use the notation in which we shall suppress the label for the dimension of spacetime for all Lagrangians that live in $6D$, and use the label only for the case of Lagrangians in $10D$.}
\be
\cL = \cL_{\rm EH} -\tfrac18 \alpha \cL_{{\rm Riem}^2} -\tfrac18 \gamma \cL_{\rm GB}\ ,
\label{2pL}
\ee
the process of going on-shell requires the elimination of the auxiliary fields using their equations of motion (EOM) following from the total Lagrangian. The EOM for the $E$-field is algebraic and it can be eliminated exactly schematically in the form $E_\m \sim V_\m + ({\rm fermi})^2$ terms\cite{Coomans:2011ih}. However, the EOM's for $V_\m$ and $Z_\m$ are not algebraic due to the curvature-squared part of the Lagrangian and therefore they become propagating. They can be eliminated order by order in the parameters $\alpha$ and $\gamma$. To first order in these parameter, the solution takes the form $V^\m= J^\m_1 + \alpha J^\m_2 +\gamma J^\m_3$ and $Z^\m = J^\m_4 + \alpha J^\m_5 + \gamma J^\m_6$, where $J^\m_1,...,J^\m_6$ are bilinear in fermionic fields. Since we shall consider actions up to quartic fermion terms and supersymmetric at first order in $\alpha$ and $\gamma$ up to cubic fermions in the supertransformations, all the auxiliary fields, namely $(V_\m, Z_\m, E_{\m\n\rh\s})$ can be set to zero. Thus, we shall consider the Lagrangian \eq{2pL} where\footnote{In taking $\cL_{GB}$ from \cite{Butter:2018wss}, their conventions can be converted to ours by letting 
$L\to e^{-2\vp},\ B_{\m\n} \to 2 B_{\m\n},\ \omega_\m{}^{ab} \to -\omega_\m{}^{ab}$ and $\ve^{\m_1...\m_6} \to -\ve^{\m_1..\m_6}$. Consequently, $H_{\m\n\rh} \to 2H_{\m\n\rh},\ \omega_{\pm \m}{}^{ab} \to -\omega_{\mp \m}{}^{ab},\  R_{\m\n}{}^{ab}(\omega) \to - R_{\m\n}{}^{ab}(\omega)$, and $R_{\m\n}{}^{ab}(\omega_\pm) \to -R_{\m\n}{}^{ab}(\omega_\mp)$.}\footnote{In getting $\cL_{{\rm Riem}^2}$ from \cite{Bergshoeff:2012ax}, we let $\cL_{EH} \to\, 2 \cL_{EH},\ L \to \, e^{-2 \vp},\ B_{\m\n} \to \, 2 B_{\m\n},\  F_{\m\n\rh}(B) \to\, 2 H_{\m\n\rh}, \ \psi_\m \to \, \sqrt2 \psi_\m,\ \vp^i \to \, - 2 e^{-2 \vp} \chi_j \del^{ij},\ \e \to \, \sqrt2 \e,\ \omega_{\pm \m}{}^{ab} \to \omega_{\pm \m}{}^{ab}$ and $R_{+\m\n}(Q) \to \sqrt2 \psi_{\m\n}$.}
\begin{align}
e^{-1} \cL_{EH} =&\, e^{- 2 \vp} \Big[ \tfrac14 R(\omega) + \pd_\m \vp \pd^\m \vp - \tfrac1{12} H_{\m\n\rh} H^{\m\n\rh} 
\nn\w2
&\, - \tfrac12 \bpsi_\m \g^{\m\n\rh} D_\n(\omega) \psi_\rh - 2 \bar{\chi} \g^{\m\n} D_\m(\omega) \psi_\n + 2 \bar{\chi} \g^\m D_\m(\omega) \chi 
\nn\w2
&\, - \tfrac1{24} H_{\m\n\rh} ( \bpsi^\s \g_{[\s} \g^{\m\n\rh} \g_{\la]} \psi^\la - 4 \bpsi_\s \g^{\s\m\n\rh} \chi - 4 \bar{\chi} \g^{\m\n\rh} \chi ) 
\nn\w2
&\, + \pd_\m \vp ( \bpsi^\m \g^\n \psi_\n - 2 \bpsi_\n \g^\m \g^\n \chi ) \Big] \ ,   
\label{Poincare}
\w2
e^{-1} \cL_{{\rm Riem}^2} =&\, R_{\m\n}{}^{ab}(\omega_-) R^{\m\n}{}_{ab}(\omega_-) + \tfrac12 \ve^{\m\n\rh\s\la\ta} B_{\m\n} R_{\rh\s}{}^{ab}(\omega_-) R_{\la\ta ab}(\omega_-) 
\nn\w2
&\, + 4 \bpsi_{ab} \g^\m D_\m(\omega, \omega_-) \psi^{ab} - 2 R_{\n\rh}{}^{ab}(\omega_-) \bpsi_{ab} \g^\m \g^{\n\rh} \psi_\m 
\nn\w2
&\, - \tfrac13 \bpsi^{ab} \g^{\m\n\rh} \psi_{ab} H_{\m\n\rh} - [ D_\m(\omega_-, \Gamma_+) R^{\m\rh ab}(\omega_-) - 4 H_{\m\n}{}^\rh R^{\m\n ab}(\omega_-) ] \bpsi_a \g_\rh \psi_b \ , 
\label{Riemann2}
\w2
e^{-1} \cL_{\rm GB} =&\, R^{\m\n\rh\s} R_{\m\n\rh\s} - 4 R^{\m\n} R_{\m\n} + R^2 
+ 2 R_{\m\n\rh\s} H^{\m\n,\rh\s} 
\nn\w2
&- 4 R^{\m\n} H^2_{\m\n} +\tfrac23 R H^2   + \tfrac{10}{3} H_4 + \tfrac19 ( H^2 )^2-2( H^2_{\m\n} )^2 
\nn\w2
& + \tfrac12 \ve^{\m\n\rh\s\ta\la} B_{\m\n} R_{\rh\s}{}^{ab}(\omega_+) R_{\ta\la ab}(\omega_+) 
+{\rm fermions} \ ,
\label{GB1}
\end{align}
where we have used the definitions
\begin{align}
\omega_{\pm \m}{}^{ab} &= \omega_\m{}^{ab} \pm  H_\m{}^{ab}\ ,
\qquad\quad H_{\m\n\rh} = 3\partial_{[\m} B_{\n\rh]}\ ,
\nn\w2
H_{\m\n,\rh\s} & := H_{\m\n\al} H_{\rh\s}{}^\al\ ,
\qquad H_4 := H_{\m\n,\rh\s} H^{\m\rh,\n\s}\ , 
\nn\w2
H^2_{\m\n} & := H_{\m \rh\s} H_\n{}^{\rh\s}\ ,
\qquad H^2 := H_{\m\n\rh} H^{\m\n\rh}\ ,
\nn\w2
\psi_{ab} 
& = 2 e_a{}^\m e_b{}^\n D_{[\mu}(\omega_+)\psi_{\n]}\ , 
\nn\w2
D_\m (\omega,\omega_-)\psi_{ab}
& = \left( \partial_\m 
+ \tfrac14 \omega_{\m pq} \gamma^{pq} \right) \psi_{ab} 
+\omega_{-\m a}{}^c \psi_{cb} + \omega_{-\m b}{}^c\psi_{ac}\ ,
\label{defs}
\end{align}
and $\Gamma_{+}= \Gamma+H$ with $\Gamma$ representing the Christoffel symbol. The fermionic part of $\cL_{GB}$ is very complicated and it can be extracted from the equations provided in \cite{Butter:2018wss}. 
Sometimes we shall use the notation $\omega_{\pm\m}{}^{ab} (H)$ to emphasize that the torsion shift is given by $H$. The Lagrangian $\cL_{{\rm Riem}^2}$ is not to be confused with another four-derivative Lagrangian that has similar form but is distinct from it. This distinct Lagrangian will be discussed in Section 2.2, and we shall refer to it as $\cL_{BdR}$ in view of the fact that it has the same form as its counterpart in $10D$ constructed long ago in \cite{Bergshoeff:1989de}. In particular note that the Einstein-Hilbert term and the four-derivative terms come with a different overall dilaton factor, unlike the BdR action in $6D$. It may seem that the two invariants are related by a duality transformation. However, this is not the case, since, as we shall see, the duality transformation must involve a combination of \eq{Riemann2} and \eq{GB1}~\cite{Liu:2013dna}.

As for the supertransformations, setting the auxiliary field to zero as explained above, they do not pick up any order $\alpha$ and $\gamma$ modifications, thereby maintaining their simple form given by 
\begin{align}
\delta e_\mu{}^a =& \bar{\epsilon} \gamma^a \psi_\mu\ ,
\nn\w2
\delta \psi_\mu =& D_\mu(\omega_+) \epsilon = D_\m(\omega)\epsilon  + \tfrac14 H_{\m\n\rh} \gamma^{\nu\rh} \e\ , 
\nn\w2
\delta B_{\mu\nu} 
=& - \bar{\epsilon} \gamma_{[\mu} \psi_{\nu]} \ , 
\nn\w2
\delta \chi
=& \tfrac{1}{2} \gamma^\mu \e \partial_\mu \vp  
+ \tfrac{1}{12} H_{\mu\nu\rho} \gamma^{\mu\nu\rho} \e
\nn\w2
\delta \varphi =&  \bar{\epsilon} \chi \ .
\label{susyAG}
\end{align}
In their  off-shell supersymmetric versions, $\cL_{EH}, \cL_{\rm Riem^2}$ and $\cL_{GB}$ are separately invariant under the off-shell supertransformations. Upon going on-shell , however, it is the sum of the Einstein-Hilbert and the $(\alpha,\gamma)$ dependent actions that is invariant under supertransformations up to first order in these arbitrary parameters. 

In summary, the bosonic part of the general two parameter Lagrangian \eq{2pL} takes the form
\be
\boxed{\begin{aligned}
\cL_{\alpha,\gamma} & =  \cL_{{\rm EH}} -\tfrac18 \alpha \cL_{{\rm Riem}^2} -\tfrac18 \gamma \cL_{{\rm GB}}  
\w2
& =  e e^{-2\vp} \Big[\, \tfrac14 R + \pd_\m \vp \pd^\m \vp - \tfrac1{12} H^{\m\n\rh}H_{\m\n\rh} \Big] 
\w2
& -\tfrac18 e \alpha \Big[\, R^{\m\n ab}(\omega_-) R_{\m\n ab}(\omega_-) +\tfrac12 \ve^{\m\n\rh\s\la\ta} B_{\m\n} R_{\rh\s}{}^{ab}(\omega_-) R_{\la\ta ab}(\omega_-) \Big] \w2
& -\tfrac18 e \gamma \Big[\, R^{\m\n\rh\s} R_{\m\n\rh\s} - 4 R^{\m\n} R_{\m\n} + R^2 
+ 2 R_{\m\n\rh\s} H^{\m\n,\rh\s} - 4 R^{\m\n} H^2_{\m\n} +\tfrac23 R H^2 
\w2
& + \tfrac{10}{3} H_4 + \tfrac19 ( H^2 )^2-2( H^2_{\m\n} )^2 + \tfrac12 \ve^{\m\n\rh\s\la\ta} B_{\m\n} R_{\rh\s}{}^{ab}(\omega_+) R_{\la\ta ab}(\omega_+) \Big] \ . 
\label{LAG}
\end{aligned}}
\ee
In studying the relation of this Lagrangian  to the type IIA action on $K3$ for $\alpha=\gamma$, it is useful to express it as~\cite{Liu:2013dna}
\begin{align}
\cL_{\alpha,\gamma} =&   e e^{-2\vp} \Big[\, \tfrac14 R + \pd_\m \vp \pd^\m \vp - \tfrac1{12} H^{\m\n\rh}H_{\m\n\rh} \Big] 
\nn\w2
& -\tfrac18 e \alpha \Big[\, R^{\m\n ab}(\omega_-) R_{\m\n ab}(\omega_-) +\tfrac12 \ve^{\m\n\rh\s\la\ta} B_{\m\n} R_{\rh\s}{}^{ab}(\omega_-) R_{\la\ta ab}(\omega_-) \Big] 
\nn\w2
& -\tfrac18 e \gamma \Big[\, - \tfrac18 \e_6\e_6 R(\omega_-)^2 - \tfrac23 \e_6\e_6 H^2 R(\omega_-) - \tfrac29 \e_6\e_6 H_4 + \tfrac12 \ve^{\m\n\rh\s\la\ta} B_{\m\n} R_{\rh\s}{}^{ab}(\omega_+) R_{\la\ta ab}(\omega_+) \Big] \ , \label{LAG2}
\end{align}
where
\begin{align}
\e_6\e_6 R(\omega_-)^2 :=&  \e^{\al\beta \mu\nu\rh\s} \e_{\al\beta abcd} R_{\m\n}{}^{ab}(\omega_-)R_{\rh\s}{}^{cd}(\omega_-)\ ,
\nn\w2
\e_6\e_6 H^2 R(\omega_-) :=& \e_{\al\mu_0...\m_4} \e^{\al \n_0...\n_4} H^{\m_1\m_2}{}_{\n_0} H_{\n_1\n_2}{}^{\m_0} R^{\m_3\m_4}{}_{\n_3\n_4} (\omega_-)\ ,
\nn\w2
\e_6\e_6 H_4 :=&  \e^{\al\beta \mu\nu\rh\s} \e_{\al\beta}{}^{abcd} H_{\m\n, ab} H_{\rh\s,cd}\ .
\end{align}
For later purposes, it is convenient to write out the $H$-dependent terms explicitly. Given that
\begin{align}
R_{\m\n ab}(\om_-) =&  R_{\m\n ab}(\om) -2D_{[\m}(\om) H_{\n] ab} +2H_{[\m}{}^{ac} H_{\n] cb}\ ,
\label{RH}\w2
\omega_{\m\n\rh}^L(\om_-) =& \omega_{\m\n\rh}^L (\om) + \Big( H_\m{}^{ab} D_\n(\om) H_{\rh ba} -\partial_\m \big( H_\n{}^{ab} \omega_{\rh ba} \big)
\nn\w2
& -H_\m{}^{ab} R_{\n\rh ba} (\om) -\tfrac23 H_\m{}^{ac} H_{\n c}{}^b H_{\rh ba} \Big)_{[\m\n\rh]}\ ,
\label{LCSH}
\end{align}
the Riemann-squared and Gauss-Bonnet actions can be written as 
\begin{align}
I({\rm Riem}^2) =& \int d^6 x\, e \Big[ R^{\mu\nu\rho\sigma} R_{\mu\nu\rho\sigma} -2 R^{\mu\nu\rho\sigma} H_{\mu\nu,\rho\sigma} +2 H^{2 \mu\nu} H^2_{\mu\nu} -2H_4
\nn\w2
& +4 \left( D_\mu H_{\nu\rho\sigma}\right)  \left(D^\nu H^{\mu\rho\sigma}\right) -4 \tH^{\m\n\rh} \Big( \omega^L_{\m\n\rh}(\om)  
-H_\m{}^{ab} D_\n (\om) H_{\rh ab}
\nn\w2
& +H_{\m\n}{}^\s R_{\rh\s} -\tfrac13 H_{\m\n}{}^\s H^2_{\rh\s}\Big)\Big] \ ,
\label{AR2}
\w2
I_{GB} =& \int d^6 x\, e \Big[R^{\m\n\rh\s} R_{\m\n\rh\s} - 4 R^{\m\n} R_{\m\n} + R^2 
+ 2 R_{\m\n\rh\s} H^{\m\n,\rh\s} 
\nn\w2
& - 4 R^{\m\n} H^2_{\m\n} +\tfrac23 R H^2   + \tfrac{10}{3} H_4 + \tfrac19 ( H^2 )^2-2( H^2_{\m\n} )^2 
\nn\w2
& -4 \tH^{\m\n\rh} \Big( \omega^L_{\m\n\rh}(\om) 
-H_\m{}^{ab} D_\n (\om) H_{\rh ab} -H_{\m\n}{}^\s R_{\rh\s} +\tfrac13 H_{\m\n}{}^\s H^2_{\rh\s}  \Big) \Big]\ ,
\label{AGB}
\end{align}
where we have used the identities \eq{id1} and \eq{id2},  we have defined
\begin{align}
\omega^L_{\m\n\rh} (\omega_\pm) =& \tr \left(
\omega_{\pm [\m} \partial_\n \omega_{\pm\rh]} + \tfrac23 \omega_{\pm[\m}\omega_{\pm\n} \omega_{\pm\rh]} \right)\ ,
\nn\w2
\tH^{\m\n\rh} =& \tfrac{1}{3!} \ve^{\m\n\rh\s\la\ta} H_{\s\la\ta}\ . 
\label{LCS}
\end{align}

\subsection*{$\cL_{\rm Riem^2}$ and $\cL_{GB}$ are not equivalent on-shell}
%%%%%%%%%%%%%%%%%%%%%%%%%%%%%%%%%%%%%%%%%%%%%%%%%%%%%%%%%%%%%%%%%%%%%%%%%%%%%%

Prior to dualization of the Lagrangian  $\cL_{\alpha,\gamma}$, it is useful to check that the $\alpha$ and $\gamma$ dependent parts are not equivalent to each other upon field redefinitions. Since the ${\rm Riem}^2$ terms cannot be changed by field redefinitions at the four-derivative order, we take $\gamma=-\alpha $ to remove these terms. From \eq{LAG}, \eq{AR2} and \eq{AGB}, we find that 
\begin{align}
\cL({\rm Riem^2}) -\cL_{GB} =& -\tfrac18 e\alpha \Big[4 R^{\m\n} R_{\m\n} -R^2 -4 R_{\m\n\rh\s} H^{\m\n,\rh\s}
+4 R^{\m\n} H^2_{\m\n} - \tfrac23 R H^2
\nn\w2
& +4 (D_\m H_{\n\rh\s})D^\n H^{\m \rh\s} + 4 ( H^2_{\m\n} )^2
- \tfrac19 ( H^2 )^2 -\tfrac{16}{3} H_4 
\nn\w2
& -8\tH^{\m\n\rh} \big( H_{\m\n}{}^\al R_{\rh\al} -\tfrac13 H^2_{\m\al} H_{\n\rh}{}^\al \big)\Big]\ .
\label{DL2}
\end{align}
To determine which terms can be removed by field redefinitions, it is convenient to use the following relations, that follow from the two-derivative Lagrangian, modulo total derivatives,
\begin{align}
& R_{\m\n} =\, -2 \vp_{\m\n} + H^2_{\m\n} + 4 \E_{\m\n} - \E_\vp g_{\m\n}\ , 
\nn\w2
& \vp^\m{}_\m =\, 2 \vp^2 - \tfrac13 H^2 - 2 \E^\m{}_\m + 2 \E_\vp \ , 
\nn\w2
&  D_\rh H^{\m\n\rh} = 2\vp_\rh H^{\m\n\rh} +e^{2\vp} \E_B^{\m\nu}\ ,
\nn\w2
& \vp^{\m\n} \vp_{\m\n} =\, - H^2_{\m\n} \vp^\m \vp^\n - \vp^2 H^2 + 2 ( \vp^2 )^2 + \tfrac19 ( H^2 )^2 
\nn\w2
&\qquad - 4 \E_{\m\n} \vp^\m \vp^\n - 6 \vp^2 \E^\m{}_\m + \tfrac43 H^2 \E^\m{}_\m + 7 \vp^2 \E_\vp - \tfrac43 H^2 \E_\vp 
\nn\w2
&\qquad + 4 ( \E^\m{}_\m )^2 + 4 \E_\vp^2 - 8 \E^\m{}_\m \E_\vp\ ,
\nn\w2
& H^2_{\m\n} \vp^{\m\n} =\, - 2 H^2_{\m\n} \vp^\m \vp^\n + \tfrac13 \vp^2 H^2 - \tfrac1{18} ( H^2 )^2 
\nn\w2
&\qquad - \tfrac13 H^2 \E^\m{}_\m + \tfrac13 H^2 \E_\vp - e^{2 \vp} H^{\m\n\rh} \vp_\m \E^B_{\n\rh}\ ,
\nn\w2
& \tH^{\m\n\rh} H_{\m\n}{}^\al \vp_{\rh\al} = \tH^{\m\n\rh}  \big( -2H_{\m\n}{}^\al \vp_\al\vp_\rh -e^{2\vp} \E_{\m\n}^B \vp_\rh \big)\ , 
\nn\w4
& (D_\s H_{\m\n\rh} ) ( D^\s H^{\m\n\rh} ) =\, 3 R^{\m\n\rh\s} H_{\m\n, \rh\s} + 2 \vp^2 H^2 - \tfrac13 ( H^2 )^2 - 3 ( H^2_{\m\n} )^2 
\nn\w2
&\qquad\qquad  - 12 H^2_{\m\n} \E^{\m\n} - 2 H^2 \E^\m{}_\m + 5 H^2 \E_\vp + 6 e^{2 \vp} H^{\m\n\rh} \vp_\m \E^B_{\n\rh} + 3 e^{4 \vp} ( \E^B_{\m\n} )^2\ ,
\label{lemmas}
\end{align}
where we have defined
\begin{align}
\E_{\m\n} &:= e^{-1} e^{2\vp} \frac{\delta\cL}{\delta g^{\m\n}}\ ,\qquad \E_\vp= e^{-1} e^{2\vp} \frac{\delta\cL}{\delta \vp}\ ,\qquad \E_{\m\n}^B= 2 e^{-1}  \frac{\delta\cL}{\delta B^{\m\n}}\ ,
\nn\w2
\vp_\mu &:= \partial_\m \vp\ ,\qquad \vp_{\m\n} := D_\m \vp_\n\ ,\qquad \vp^2 := \vp^\m \vp_\m\ .
\end{align}
Using the identities \eq{lemmas}, the Lagrangian \eq{DL2} becomes
\begin{align}
\cL({\rm Riem^2}) -\cL_{GB}=&\,  -\tfrac18 e\, \alpha \Big[ 8 ( H^2_{\m\n} )^2 - \tfrac43 ( H^2 )^2 - \tfrac{16}3 H_4 + 32 H^2_{\m\n} \vp^\m \vp^\n - \tfrac{16}3 \vp^2 H^2 + 16 ( \vp^2 )^2 
\nn\w2
& -8 \tH^{\m\n\rh} \big( 4 H_{\m\n}{}^\al \vp_\rh \vp_\al  +\tfrac23 H^2_{\m\al} H_{\n\rh}{}^\al \big)
& \nn\w2
& -8 \tH^{\m\n\rh}  \big( 4H_{\m\n}{}^\al \E_{\rh\al} +2 e^{2\vp} \E^B_{\m\n} \vp_\rh \big) + 32 H^2_{\m\n} \E^{\m\n}
\nn\w2
&\,  - 64 \E_{\m\n} \vp^\m \vp^\n - 64 \E_{\m\n} \vp^{\m\n} - \tfrac{16}3 H^2 \E^\m{}_\m 
\nn\w2
&\, - 32 \vp^2 \E^\m{}_\m + 64 \vp^2 \E_\vp + 32 e^{2 \vp} H^{\m\n\rh} \vp_\m \E^B_{\n\rh} 
\nn\w2
&\, + 64 ( \E_{\m\n} )^2 - 32 \E_\vp \E^\m{}_\m + 20 \E_\vp^2 + 4 e^{4 \vp} ( \E^B_{\m\n} )^2 \Big] 
\end{align}
All the terms involving factors of the lowest order equations of motion can clearly be removed by field redefinitions in the two-derivative part of the Lagrangian. Then we are left with terms which schematically have form $H^4, H^2 (\partial\vp)^2$ and $(\partial\vp)^4$. Thus, we see that the Gauss-Bonnet Lagrangian $\cL_{GB}$ cannot be brought into the form of the Riemann-squared Lagrangian $\cL({\rm Riem}^2)$ by field redefinitions. 

\subsection{The Lagrangian $\cL_{BdR}$ in $6D$ }
%%%%%%%%%%%%%%%%%%%%%%%%%%%%%%%%%%%%%%%%%%%%%%%%%%%%%%%%%

The $6D$ analog of the $10D$ Bergshoeff-de Roo Lagrangian, including fermionic terms (up to quartics) was obtained in \cite{Chang:2021tsj} by dimensional reduction of the former on $T^4$ followed by a consistent truncation. The bosonic part of the Lagrangian is given by 
\begin{align}
\cL_{BdR}  =& ee^{2\vp} \Bigg[
 \tfrac14 R +  g^{\m\n} \partial_\m \vp \partial_\n \vp - \tfrac{1}{12} G_{\m\n\rh} G^{\m\n\rh}
\nn\w2
& +\alpha \Big(
\,G^{\m\n\rh} \omega^L_{\m\n\rh}(\omega_-)   - \tfrac14 R_{\m\n rs}(\omega_-) R^{\m\n rs}(\omega_-) \Big)\,\Bigg]\ ,
\label{BdR4}
\end{align}
where
\be
\omega_{\pm \m ab} = \omega_{\m ab} \pm G_{\m ab}\ ,\qquad  G_{\m\n\rh} = 3\partial_{[\m} C_{\n\rh]}\ .
\ee
As is well known, the term $G^{\m\n\rh} \omega^L_{\m\n\rh}(\omega_-)$ can be absorbed into the definition of $G=dC$ to define
\be
\cG_{\m\n\rh} = 3 \partial_{[\m} C_{\n\rh]} -6\alpha\, \omega^L_{\m\n\rh}(\omega_-)\ .
\label{cG}
\ee
and the local Lorentz invariance of this field strength requires that $C_{\m\n}$ transforms as
\be
\delta_\Lambda C_{\m\n} = 2 \alpha \tr (\omega_{-[\m} \partial_{\n]} \Lambda)\ .
\label{LL}
\ee
For later purposes, let us spell out the torsion dependence in $\cL_{BdR}$. Using \eq{RH}  and \eq{LCSH} (with $H$ replaced by $G$) and the following lemma
\begin{align}
& e e^{2\varphi} G^{\mu\nu\rho} \omega^L_{\mu\nu\rho}(\omega_-)
= e e^{2\varphi} \Bigl[ G^{\mu\nu\rho} \omega^L_{\mu\nu\rho}(\omega) 
+ R^{\mu\nu\rho\sigma}(\omega) G_{\m\n,\rh\s}-\tfrac{2}{3} G^{\m\n,\rh\s} G_{\m\rh,\n\s}
-G^{\m\n\rh} \partial_\m\left( \omega_\n{}^{rs} G_{\rh rs}\right) \Big]\ ,
\label{r2}
\end{align}
we obtain
\begin{align}
\cL_{BdR}  =& ee^{2\vp} \Bigg[
 \tfrac14 R +  g^{\m\n} \partial_\m \vp \partial_\n \vp - \tfrac{1}{12} G_{\m\n\rh} G^{\m\n\rh}
\nn\w2
& +\alpha\Big(
\,G^{\m\n\rh} \omega^L_{\m\n\rh}(\omega)   - \tfrac14 R_{\m\n\rh\s} R^{\m\n\rh\s} +\tfrac32 R_{\m\n\rh\s} G^{\m\n,\rh\s} -\tfrac16 G_4
\nn\w2
& -\tfrac12 G^2_{\m\n} G^{2\m\n} - \left(D_\m G_{\n\rh\s} \right) D^\n G^{\m\rh\s} -G^{\m\n\rh} \partial_\m \left(\omega_\n{}^{ab} G_{\rh ab} \right)\Big)\,\Bigg]\ .
\label{BdR5}
\end{align}
where the definitions \eq{defs} with $H$ replaced by $G$ have been used. 

%%%%%%%%%%%%%%%%%%%%%%%%%%%%%%%%%%%%%%%%%%%%%%%%%%%%%%%%%%%%%%%%%%%%%%%%%%
\section{Dualization of the Lagrangian $\cL_{\alpha,\gamma}$ in $6D$ }
%%%%%%%%%%%%%%%%%%%%%%%%%%%%%%%%%%%%%%%%%%%%%%%%%%%%%%%%%%%%%%%%%%%%%%%%%%

In this section, we shall formulate the Lagrangian $\cL_{\al,\gamma}$ such that the field equations and Bianchi identities associated with the two-form potential $B$ are interchanged. As is well known, in $6D$ this involves a dual two-form potential which we shall denote by $C$. In one approach, one can work at the level of field equations and Bianchi identities, and after performing the dualization map, try to find the ``dual Lagrangian'' which will produce this system.  This procedure can get quite cumbersome. An alternative method which works at the level of the Lagrangian throughout, thereby not requiring the integrating-out process from a set of field equations, is what we refer to as the Hodge-dualization method. This is a well known method in which one adds a total Lagrange multiplier term of the form $H \wedge dC$ where $H=dB$ and integrate over $H$ which is treated as independent variable. For this approach to work, the $B$-filed should arise through its field strength $H$ everywhere in the action. This requirement holds in the Lagrangian $\cL_{\al,\gamma}$ we are considering. 

\subsection{The Lagrange multiplier method}
%%%%%%%%%%%%%%%%%%%%%%%%%%%%%%%%%%%%%%%%%%%%%%%%

Adding to the Lagrangian $\cL_{\al,\gamma}$ given in \eq{LAG} a total derivative Lagrange multiplier term
\be
\Delta \cL (B,C) =  \tfrac{1}{2\times 3!} \epsilon^{\m\n\rh\s\ta\la} H_{\m\n\rh}\partial_\s C_{\ta\la}\ ,
\ee
where $C_{\m\n}$ is the dual potential, and using \eq{RH} we have 
\begin{align}
& \cL_{\al,\gamma}+\Delta \cL(B,C) =  e e^{-2\vp} \Big[\, \tfrac14 R + \pd_\m \vp \pd^\m \vp - \tfrac1{12} H^{\m\n\rh} H_{\m\n\rh}\Big]   
\nn\\
&\qquad  -\tfrac18 \alpha e \Big[\, R^{\m\n\rh\s} R_{\m\n\rh\s} +4 D_\m H_{\n ab} D^\n H^{\m ab} -2 R_{\m\n\rh\s} H^{\m\n,\rh\s}
+2H^2_{\m\n} H^{2\m\n} -2 H_4 \Big] 
\nn\\
&\qquad -\tfrac18 \gamma e \Big[\, R^{\m\n\rh\s} R_{\m\n\rh\s} - 4 R^{\m\n} R_{\m\n} + R^2 
+ 2 R_{\m\n\rh\s} H^{\m\n,\rh\s} - 4 R^{\m\n} H^2_{\m\n} + \tfrac23 R H^2
\nn\\
&\qquad  + \tfrac{10}{3} H_4  - 2 ( H^2_{\m\n} )^2 + \tfrac19 ( H^2 )^2
\Big] + \tfrac{1}{36} \epsilon^{\m\n\rh\s\ta\la} H_{\m\n\rh}\cG_{\s\tau\lambda}   \ ,
\end{align}
where the ${\widetilde H}^{\m\n\rh} \omega^L_{\m\n\rh}(\omega_\pm)$ terms present in the Lagrangian $\cL_{\al,\gamma}$ have been absorbed into the Lagrange multiplier term now involving the field strength 
\bea
\cG_{\m\n\rh} &=& G_{\m\n\rh} -3 \alpha\, \omega^L_{\m\n\rh}(\omega_-(H)) -3 \gamma\, \omega^L_{\m\n\rh}(\omega_+(H))
\nn\w2
G_{\m\n\rh} &=& 3\partial_{[\m} C_{\n\rh]}\ ,
\nn\w2
\tG^{\m\n\rh} &=& \tfrac{1}{3!} \ve^{\m\n\rh\s\la\ta} G_{\s\la\ta}\ ,\qquad \mbox{idem}\quad  \tcG^{\m\n\rh}\ .
\label{tGdef}
\eea
Note that the Lorentz Chern-Simons form depend on spin connection shifted by torsion given by $H$ instead of $\cG$. It is convenient to group the  terms in this Lagrangian as follows 
\begin{align}
& \cL_{\al,\gamma}+\Delta \cL(B,C) =  \cL_{01} + \cL_1\ 
\nn\w2
&\qquad e^{-1} \cL_{01} =   e^{-2\vp} \Big[\, \tfrac14 R + \pd_\m \vp \pd^\m \vp - \tfrac1{12} H^{\m\n\rh} H_{\m\n\rh} \Big]  + \tfrac16 H^{\m\n\rh} \tcG_{\m\n\rh} \ , 
\nn\w2
& \qquad e^{-1} \cL_1 \  = -\tfrac18 \Big[ 
(\alpha+\gamma)R^{\m\n\rh\s} R_{\m\n\rh\s}- 4\gamma R^{\m\n} R_{\m\n} + \gamma R^2
+2(-\alpha+\gamma) R_{\m\n\rh\s} H^{\m\n,\rh\s}
\nn\w2
& \qquad - 4\gamma R^{\m\n} H^2_{\m\n} + \tfrac23 \gamma R H^2 +4\alpha ( D_\m H_{\n\rh\s} ) D^\n H^{\m \rh\s}  +2(\alpha-\gamma) ( H^2_{\m\n} )^2
\nn\w2
& \qquad  + \tfrac19 \gamma ( H^2 )^2 +\left(-2\alpha+\tfrac{10}{3}\gamma\right) H_4 \Big]\ .
\label{DL1}
\end{align}
The first term is labelled as $\cL_{01}$ to denote the fact that it contains up to and including terms that are first order in parameters $\alpha$ and $\gamma$. Treating $H_{\m\n\rh}$ as an independent field, its field equation is 
\begin{align}
& \tcG^{\m\n\rh}   =  e^{-2\vp} H^{\m\n\rh} -3 \tH^{\alpha\beta\gamma} \frac{\delta}{\delta H_{\m\n\rh}} \Big( \alpha\,\omega^L_{\alpha\beta\gamma}(\omega_-(H)) + \gamma\,\omega^L_{\alpha\beta\gamma}(\omega_+(H)) \Big)     
-6e^{-1} \frac{\delta \cL_1}{\delta H_{\m\n\rh}} \ .
\label{AGDE}
\end{align}
This is a nonlinear equation in $H$.  For our purposes here, we need the solution for $H$ in terms of $G$ only to first order in $\alpha$ and $\gamma$  which is readily seen to be 
\be
\boxed{
\begin{aligned}
H^{\m\n\rh} =& e^{2\vp}\Bigg\{  \tcG^{\m\n\rh} +3 \tH^{\alpha\beta\gamma} \frac{\delta}{\delta H_{\m\n\rh}} \Big[ \alpha\,\omega^L_{\alpha\beta\gamma}(\omega_-(H)) + \gamma\,\omega^L_{\alpha\beta\gamma}(\omega_+(H)) \Big]  + 6 e^{-1} \frac{\delta \cL_1}{\delta H_{\m\n\rh}} \,\Bigg\}\Bigg|_{H=e^{2\vp} \tG}\ ,
\label{de1}
\end{aligned}}
\ee
where we recall that $\omega_\pm(H) = \omega \pm H$. Writing $H=e^{2\vp} \tG + \alpha H_1 +\gamma H_2$, without having to specify $H_1$ and $H_2$, it is easy to verify that up to order $\alpha$ and $\gamma$ the last two terms in $\cL_{01}$ simplify as 
\begin{align}
\Big( e^{-2\vp} H_{\m\n\rh} & H^{\m\n\rh} -2 \tcG^{\m\n\rh} H_{\m\n\rh}\Big)\Big|_{H=e^{2\vp} \tG} = e^{2\vp} \cG^{\m\n\rh}\cG_{\m\n\rh}\Big|_{H=e^{2\vp} \tG}\ . 
\end{align}
Thus the dualized action to first order in $\alpha$ and $\gamma$ is given by
\begin{align} 
e^{-1} \cL_{\al,\gamma}^{\rm dual}  =&  \Big[  e^{-2\vp} \Big(\, \tfrac14 R + \pd_\m \vp \pd^\m \vp - \tfrac1{12} e^{4\vp} \cG^{\m\n\rh} \cG_{\m\n\rh} \Big)  +\cL_1 \Big]_{H=e^{2\vp} \tG}\ .
\end{align}
Working out the last two terms explicitly in terms of $G$ we find, to first order in $\alpha$ and $\gamma$,
\begin{align}
& -\tfrac{1}{12} e e^{2\vp}  \cG^{\m\n\rh}\cG_{\m\n\rh}\Big|_{H=e^{2\vp} \tG}  
\nn\w2
& = e e^{2\vp} \Big[-\tfrac{1}{12} G_{\m\n\rh}G^{\m\n\rh} +\tfrac12 (\alpha+\gamma) G^{\m\n\rh} \omega^L_{\m\n\rh}(\omega) -\tfrac12 ( \alpha - \gamma ) G^{\m\n\rh} \pd_\m ( \omega_\n{}^{ab} H_{\rh ab}) \Big]
\nn\w2
& \ \ -\tfrac16 (\alpha-\gamma) \tG^{\m\n\rh} \big(  e^{8\vp} G^2_{\m\alpha} G_{\n\rh}{}^\alpha-3 e^{4\vp} G_{\m\n}{}^\alpha R_{\rh\alpha} \big) 
\end{align}
\begin{align}
e^{-1} \cL_1 \Big|_{H=e^{2\vp} \tG} =& -\tfrac18 (\alpha+\gamma) R_{\m\n\rh\s}R^{\m\n\rh\s} -\tfrac18 \gamma(-4R_{\m\n}R^{\m\n} +R^2) 
\nn\w2
& -\tfrac18 e^{8\vp} \Big[\tfrac23(-3\alpha+5\gamma)  G^{\m\n,\rh\s} G_{\m\rh,\n\s} + 2(\alpha-\gamma) G^2_{\m\n}G^{2\m\n} +\tfrac19 \gamma\left(G^2\right)^2\Big]
\nn\w2
&
-\tfrac18 e^{4\vp}\Big[2(\alpha-\gamma)  R_{\m\n\rh\s} G^{\m\n,\rh\s} -4\alpha R^{\m\n} G^2_{\m\n} +\tfrac23 \alpha R G^2 
\nn\w2
&-\tfrac43 \alpha (D_\m G_{\n\rh\s} )(D^\m G^{\n\rh\s}) +4\alpha (D^\rh G_{\rh\m\n})(D_\ta G^{\ta\m\n}) -\tfrac{16}{3}\alpha G^{\m\n\rh} \vp^\s D_\s G_{\m\n\rh} 
\nn\w2
& +16 \alpha \vp^\rh G_{\rh\m\n} D_\s G^{\s\m\n} +16\alpha G^2_{\m\n} \vp^\m \vp^\n -\tfrac{16}{3}\alpha G^2\vp^2 \Big]\ ,
\end{align}
where we have used the identities \eq{id1} and \eq{id2} with $H$ replaced by $G$. 
 Adding the  two contributions to the Lagrangian, the resulting  $\cL= \cL_{01}+ \cL_1$, prior to rescaling of the metric and prior to identification of EOM terms, is given by 
\begin{align}
e^{-1}\cL_{\al,\gamma}^{\rm dual} =&  e^{-2\vp} \Big[ \tfrac14 R + \pd_\m \vp \pd^\m \vp - \tfrac1{12} e^{4\vp} G^{\m\n\rh} G_{\m\n\rh} 
-\tfrac12 ( \alpha - \gamma ) e^{4 \vp} G^{\m\n\rh} \pd_\m ( e^{2 \vp} \om_\n{}^{ab} \tG_{\rh ab} ) \Big]
\nn\w2
& +\Big[ \tfrac12 (\alpha+\gamma) e^{2\vp} G^{\m\n\rh} \omega^L_{\m\n\rh}(\omega) -\tfrac18 (\alpha+\gamma) R_{\m\n\rh\s}R^{\m\n\rh\s} -\tfrac18 \gamma(-4R_{\m\n}R^{\m\n} +R^2) \Big]
\nn\w2
& -\tfrac18 e^{4\vp} \Big[ 2(\alpha-\gamma)R_{\m\n\rh\s} G^{\m\n,\rh\s} -4\alpha R^{\m\n} G^2_{\m\n} +\tfrac23 \alpha R G^2 
\nn\w2
&-\tfrac43 \alpha (D_\m G_{\n\rh\s} )(D^\m G^{\n\rh\s}) +4\alpha (D^\rh G_{\rh\m\n})(D_\ta G^{\ta\m\n}) -\tfrac{16}{3}\alpha G^{\m\n\rh} \vp^\s D_\s G_{\m\n\rh} 
\nn\w2
& +16 \alpha \vp^\rh G_{\rh\m\n} D_\s G^{\s\m\n} +16\alpha G^2_{\m\n} \vp^\m \vp^\n -\tfrac{16}{3}\alpha G^2\vp^2 \Big] 
\nn\w2
& -\tfrac18 e^{8\vp} \Big[\tfrac23(-3\alpha+5\gamma)  G^{\m\n,\rh\s} G_{\m\rh,\n\s} + 2(\alpha-\gamma) G^2_{\m\n}G^{2\m\n} +\tfrac19 \gamma\left(G^2\right)^2\Big]
\nn\w2
& -\tfrac16 (\alpha-\gamma) \tG^{\m\n\rh} \big(  e^{8\vp} G^2_{\m\alpha} G_{\n\rh}{}^\alpha-3 e^{4\vp}G_{\m\n}{}^\alpha R_{\rh\alpha} \big)\ .
\label{total}
\end{align}
The supertransformations of the dualized theory can be derived as follows. Those of $e_\m{}^a$ and $\vp$ in \eq{susyAG} remain the same. In the supertransformations of $\psi_\m$ and $\chi$, bearing in mind that we are considering terms up to cubic fermion terms and at first order in $\alpha$ and $\gamma$, it suffices to use the duality equation to replace $H$.
There remains the supertransformation of $C_{\m\n}$. To find it, we follow the method provided in \cite{BdR90} and we seek the cancellation of the terms that arise from the variation of $C_{\mu\nu}$ in the Lagrangian \eq{DL1} (including the fermion terms up to the quartics) with the Lagrange multiplier term, and treat $H_{\m\n\rh}$ as independent variable, thereby not using $dH=0$. Putting aside the fermionic parts of $\cL_{{\rm Riem}^2}$ and $\cL_{GB}$, the variations that contain $dH$ are 
\begin{align}
\delta \cL_{01}\Big|_{dH} = &~\tfrac{1}{12} \e^{\m\n\n_1...\n_4} \big(\partial_{\n_1} H_{\n_2..\n_4} \big)\Big(  \delta C_{\m\n} + \al \tr \big(\omega_{-\m}\delta \omega_{-\n}\big)_{H=e^{2\vp} \tG} +\gamma \tr \big(\omega_{+\m}\delta \omega_{+\n}\big)_{H=e^{2\vp} \tG} 
\nn\w2
& +e^{-2\vp} {\bar\e} \gamma_\m \psi_\n -e^{-2\vp} {\bar\e} \gamma_{\m\n} \chi \Big)\ . \
\end{align}
The last two terms come from the variation of the Pauli coupling in $\cL_{EH}$. The other source of variations that contain $dH$ come from the following lemma in which the fermionic terms are omitted:
\be
\delta \psi_{ab} = \tfrac14 \gamma^{cd} R_{abcd}(\omega_+)\,\e = \tfrac14 \gamma^{cd} R_{cdab}(\omega_-)\,\e + \gamma^{cd} \partial_{[a} H_{bcd]}\,\e\ .
\label{dpsi}
\ee
The curvature $R_{abcd}(\omega_-)$ transforms like the gaugino of the Yang-Mills multiplet, and it is the part which participates in the usual cancellations in the Noether procedure involving  $\cL_{{\rm Riem}^2}$. As to the variation of $\cL_{GB}$, it is not clear from the results given in \cite{BdR90} whether the $dH$ terms arise in that way. Therefore, we shall write the $dH$ involving terms coming from $\cL_1$ as follows
\begin{align}
\delta \cL_1\Big|_{dH} =&  \e^{\m\n\n_1...\n_4} \big(\partial_{\n_1} H_{\n_2..\n_4} \big) \Big(  -\tfrac{1}{8\times 24}\,\al\,{\bar\e} \gamma_{\m\n ab}\,\frac{\delta \cL_{{\rm Riem}^2}}{\delta \bpsi_{ab}} -\tfrac18 \gamma \cO_{\m\n} \Big) ,
\end{align}
where we have written all the terms proportional to $dH$ that arise in the variation of $\cL_{GB}$ as $ \e^{\m\n\n_1...\n_4} \big(\partial_{\n_1} H_{\n_2..\n_4}\big) \cO_{\m\n}$. Thus, requiring that all the terms containing $dH$ cancel, we find that
\begin{align}
\delta C_{\m\n}=&   e^{-2\vp}\left( -{\bar\e} \gamma_{[\m} \psi_{\n]} + {\bar\e} \gamma_{\m\n} \chi\right) -\Big[\al \tr \big(\omega_{-[\m}\delta \omega_{-\n]}\big) +\gamma \tr \big(\omega_{+[\m}\delta \omega_{+\n]}\big)
\nn\w2
& -\tfrac{1}{16}\,\al\,{\bar\e} \gamma_{\m\n ab}\,\frac{\delta \cL_{{\rm Riem}^2}}{\delta \psi_{ab}} -\tfrac32 \gamma \cO_{\m\n}\Big]_{H=e^{2\vp} \tG}\ .
\end{align}

\subsection{The string frame}
%%%%%%%%%%%%%%%%%%%%%%%%%%%%%%%%%%%

Next, we rescale the metric as
\be
g_{\m\n} \to g'_{\m\n} = e^{2\vp} g_{\m\n}\ .
\label{mrs}
\ee
As a consequence, we have
\begin{align}
{\Gamma'}_{\m\n}^\rh & = \Gamma_{\m\n}^\rh + \delta_\m^\rh \vp_\n + \delta_\n^\rh \vp_\m -g_{\m\n} \vp^\rh\ ,
\nn\w2
\omega^\prime_\m{}^{ab} =&\, \omega_\m{}^{ab} + 2 e_\m{}^{[a} \vp^{b]} \ ,
\nn\w2
D'_\s G_{\m\n\rh} & = D_\s G_{\m\n\rh} - 3 G_{\s[\n\rh} \vp_{\m]} - 3 G_{\m\n\rh} \vp_\s + 3 g_{\s[\rh} G_{\m\n] \ta} \vp^\ta \ ,
\nn\w2
R_{\m\n\rh\s} (g') & = e^{2\vp} \Big[ R_{\m\n\rh\s} -4 \Big( g_{\m\rh}\vp_{\n\s}-g_{\m\rh}\vp_\n\vp_\s +\tfrac12 g_{\m\rh}g_{\n\s} \vp^2\Big)\Big|_{[\m\n][\rh\s]}\,\Big]\ ,
\nn\w2
R_{\m\n}(g') =&  R_{\m\n} - g_{\m\n} \vp^\rh{}_\rh - 4 \vp_{\m\n} + 4 \vp_\m \vp_\n - 4 g_{\m\n} \vp^2 \ ,
\nn\w2
R (g')  =& e^{-2\vp} \big(  R - 10 \vp^\m{}_\m - 20 \vp^2 \big)\ ,
\nn\w2
\omega^{'L}_{\m\n\rh} =&\, \omega^L_{\m\n\rh}(\omega) + 2 \partial_{[\m} ( \omega_\n{}^{ab} e_{\rh] a} \vp_b )\ .
\label{hats}
\end{align}
Note in particular that $D^{\prime\la} G_{\la\m\n} = D^\la G_{\la\m\n}$. Substituting these results in the primed version of the total Lagrangian \eq{total}, we obtain one of the key results of this paper taking the form
\be
\boxed{\begin{aligned}
e^{-1}\cL_{\al,\gamma}^{\rm dual} =&\,  e^{2\vp} \Bigg[ \tfrac14 R + \pd_\m \vp \pd^\m \vp - \tfrac1{12} G^{\m\n\rh} G_{\m\n\rh} -\tfrac12 ( \alpha - \gamma ) G^{\m\n\rh} \pd_\m ( \omega_\n{}^{ab} \tG_{\rh ab} ) 
\w2
&\, +( \alpha - \gamma ) G^{\m\n\rh} \pd_\m ( \vp^\s \tG_{\n\rh\s} ) + ( \alpha + \gamma ) G^{\m\n\rh} \partial_\m ( \omega_\n{}^{ab} e_{\rh a} \vp_b ) 
\w2
& +\tfrac12 ( \alpha + \gamma ) G^{\m\n\rh} \omega^L_{\m\n\rh}(\omega) - \tfrac18 ( \alpha + \gamma ) R_{\m\n\rh\s} R^{\m\n\rh\s} -\tfrac14 \gamma( - 2 R_{\m\n} R^{\m\n} + \tfrac12 R^2 ) 
\w2
& -\tfrac14( \alpha - \gamma ) R_{\m\n\rh\s} G^{\m\n,\rh\s} +\tfrac12 \alpha R^{\m\n} G^2_{\m\n} - \tfrac{1}{12} \alpha R G^2 
\w2
& - ( \alpha - 3 \gamma ) R_{\m\n} \vp^\m \vp^\n + ( \alpha - 3 \gamma ) R_{\m\n} \vp^{\m\n} +\tfrac12 ( \alpha + 3 \gamma ) R \vp^2 +\tfrac32 \gamma R \vp^\m{}_\m 
\w2
& - ( \alpha - \gamma ) G^2_{\m\n} \vp^\m \vp^\n - ( \alpha + \gamma ) G^2_{\m\n} \vp^{\m\n} + (\tfrac56 \alpha -\tfrac12 \gamma ) G^2 \vp^2 +\tfrac13 \alpha G^2 \vp^\m{}_\m 
\w2
& -2 ( \alpha - 3 \gamma ) \vp^{\m\n} \vp_{\m\n} -\tfrac12 ( \alpha + 12 \gamma ) ( \vp^\m{}_\m )^2 +4 ( \alpha - 3 \gamma ) \vp_{\m\n} \vp^\m \vp^\n 
\w2
& -2 ( 2 \alpha + 9 \gamma ) \vp^2 \vp^\m{}_\m -5 ( \alpha + 3 \gamma ) ( \vp^2 )^2 
\w2
&+\tfrac16 \alpha (D_\m G_{\n\rh\s} )(D^\m G^{\n\rh\s}) -\tfrac12\alpha (D^\la G_{\la\m\n})(D_\ta G^{\ta\m\n}) 
\w2
& -\tfrac23 \alpha G^{\m\n\rh} \vp^\s D_\s G_{\m\n\rh} - \alpha \vp^\rh G_{\rh\m\n} D_\la G^{\la\m\n} 
\w2
& -\tfrac{1}{12}(-3\alpha+5\gamma)  G^{\m\n,\rh\s} G_{\m\rh,\n\s} -\tfrac14 (\alpha-\gamma) G^2_{\m\n}G^{2\m\n} -\tfrac{1}{72} \gamma\left(G^2\right)^2
\w2
&  -\tfrac16 (\alpha-\gamma) \tG^{\m\n\rh} \Big(G^2_{\m\alpha} G_{\n\rh}{}^\alpha -3 G_{\m\n}{}^\alpha \big( R_{\rh\alpha}-4\vp_{\rh\alpha} +4\vp_\rh\vp_\alpha\big)\Big)  \Bigg]
\label{genDL}
\end{aligned}}
\ee
Setting $\gamma=0$ gives the result of dualizing $\cL_{\rm Riem^2}$, and  setting $\alpha=0$ gives the result of dualizing $\cL_{GB}$. While several terms can be removed by field redefinitions, such a step will modify the simple supersymmetry transformations by introducing the corresponding $\alpha$ or $\gamma$ dependent higher derivative terms.

\subsection{Comparison with $\cL_{BdR}$ } 
%%%%%%%%%%%%%%%%%%%%%%%%%%%%%%%%%%%%%%%%%%%%%%%%%%%%%%

To compare $\cL_{\al,\gamma}^{\rm dual}$ with $\cL_{BdR}$ we set $\gamma=\alpha$ in \eq{genDL} and examine the difference
\begin{align}
\cL_{\al,\al}^{\rm dual}-\cL_{BdR} =&\,  \alpha e e^{2\vp} \Bigg[ 
2  G^{\m\n\rh} \partial_\m ( \omega_\n{}^{ab} e_{\rh a} \vp_b ) +G^{\m\n\rh} \partial_\m \left(\omega_\n{}^{ab} G_{\rh ab} \right)
\nn\w2
&
-\tfrac32 R_{\m\n\rh\s} G^{\m\n,\rh\s}  -\tfrac{1}{72} \left(G^2\right)^2
 +\tfrac12 G^2_{\m\n} G^{2\m\n} 
\nn\w2
&  -\tfrac14  ( - 2 R_{\m\n} R^{\m\n} + \tfrac12 R^2 ) 
+\tfrac12  R^{\m\n} G^2_{\m\n} - \tfrac{1}{12}  R G^2 
\nn\w2
& +2 R_{\m\n} \vp^\m \vp^\n -2 R_{\m\n} \vp^{\m\n} +2 R \vp^2 +\tfrac32  R \vp^\m{}_\m 
\nn\w2
& - 2 G^2_{\m\n} \vp^{\m\n} +\tfrac13  G^2 \vp^2 +\tfrac13  G^2 \vp^\m{}_\m 
\nn\w2
& +4 \vp^{\m\n} \vp_{\m\n} -\tfrac{13}{2}  ( \vp^\m{}_\m )^2 -8 \vp_{\m\n} \vp^\m \vp^\n 
-22 \vp^2 \vp^\m{}_\m -20  ( \vp^2 )^2 
\nn\w2
&+\left(D_\m G_{\n\rh\s} \right) D^\n G^{\m\rh\s} +\tfrac16  (D_\m G_{\n\rh\s} )(D^\m G^{\n\rh\s}) -\tfrac12 (D^\la G_{\la\m\n})(D_\ta G^{\ta\m\n}) 
\nn\w2
& -\tfrac23  G^{\m\n\rh} \vp^\s D_\s G_{\m\n\rh} -  \vp^\rh G_{\rh\m\n} D_\la G^{\la\m\n} 
 \Bigg]
\end{align}
Next, we show that this difference can be removed entirely by field redefinitions. To this end we use the field equations obtained from the 2-derivative part of the Lagrangian \eq{genDL} through
\be
\E_{\m\n} := e^{-1} e^{-2\vp} \frac{\delta\cL}{\delta g^{\m\n}}\ ,\qquad \E_\vp=  e^{-1} e^{-2\vp} \frac{\delta\cL}{\delta \vp}\ ,\qquad \E_{\m\n}^C= 2 e^{-1}  \frac{\delta\cL}{\delta C^{\m\n}}\ ,
\ee
and computing the analogs of \eq{lemmas} for this case, we find
\be
\boxed{\begin{aligned}
\cL^{\rm dual}_{\al,\al} -\cL_{BdR} &= \alpha e \Big[ - 2 \omega_\m{}^{ab} e_{\n a} \vp_b  - \omega_\m{}^{ab} G_{\n ab} + 2 G_{\m\n\rh} \vp^\rh \Big] \E^{C \m\n}
\w2
&\, + \alpha\, e\, e^{2 \vp} \Big[ 8 \E_{\m\n} \E^{\m\n} + 2 \E_\vp \E^\m{}_\m - 2 ( \E^\m{}_\m )^2 + \tfrac52 \E_\vp^2 + e^{- 4 \vp} ( \E^C_{\m\n} )^2 \Big] 
\end{aligned} }
\ee
which shows that $\cL^{\rm dual}_{\al,\al}$ equals $\cL_{BdR}$ upon  performing the field redefinitions\footnote{The terms that involve the square of the EOM's can be removed in two different ways, since given a term of the form $f \E_1\E_2$ where $f$ is constant or field dependent term, and $\E_1, \E_2$ refer to the EOM of fields $\vp_1$ and $\vp_2$, we can either make the field redefinition $\delta\vp_1= -f\E_2$ or $\delta \vp_2 =-f\E_1$. Here we have chosen one of these two ways.}
\begin{align}
\delta C_{\m\n} =&  -2\alpha \big(  2 \omega_{[\m}{}^{ab} e_{\n] a} \vp_b + \omega_{[\m}{}^{ab} G_{\n] ab}  - 2 G_{\m\n\rh} \vp^\rh -e^{-2\vp} \E_{\m\n}^C \big)\ ,
\nn\w2
\delta g_{\m\n} =& \alpha \left( 8\E_{\m\n} -2 g_{\m\n}  \E^\la{}_\la + 2  g_{\m\n} \E_\vp\right) \ ,
\nn\w2
\delta \vp =& \tfrac52 \alpha \E_\vp\ .
\label{rc}
\end{align}
For completeness, it is useful to also determine the local gauge transformations that leave the dual Lagrangian invariant. Those associated with the metric and dilaton remain the same, but the Local Lorentz and supersymmetry transformations of $C_{\m\n}$ get deformed. To begin with, the local Lorentz invariance of the duality equation \eq{de1} implies that $\cG_{\m\n\rh}$ is invariant. From \eq{hats}, and taking into account the rescaling of the metric, it follows that under local Lorentz transformations $C_{\m\n}$ acquires the transformation
 \be
 \delta_\Lambda C_{\m\n} = 2\alpha \tr (\omega_{[\m} \partial_{\n]} \Lambda) +4\alpha \vp^a e_{[\m}{}^b  \partial_{\n]}\Lambda_{ab} \ .
 \ee
Consequently, the redefined $C^\prime_{\m\n}= C_{\m\n} + \delta C_{\m\n}$, upon using \eq{rc}, and recalling the redefinition of the metric, transforms as 
\be
\delta_\Lambda C^\prime_{\m\n} = 2\alpha \tr (\omega_{[\m} \partial_{\n]} \Lambda) + 2\alpha G_{[\m}{}^{ab} \partial_{\n]}\Lambda_{ab} = 2\alpha \tr (\omega_{-[\m} \partial_{\n]} \Lambda)\ ,
\ee
in agreement with \eq{LL}.
%
%HERE WE HAVE DROPPED A WHOLE SECTION ON THE DUALIZATION OF The 6D BdR LAGRANGIAN, AND SIMPLY GIVE The BOTTOM LINE  RESULT. THIS DROPPED SECTION CAN BE FOUND IN GB5.tex
%
After a long and tedious calculation we have also established the following result for
the dual of  $\cL_{BdR}^{\rm dual}$: 
\begin{align}
\cL_{BdR}^{\rm dual} =& 
 \cL_{\al,\al}-\tfrac14 \al\, e\, \Big[ 4 \tH^{\m\n\rh} H_{\n\rh}{}^\al \E^B_{\m\al} -6 e^{4 \vp} \E^B_{\m\n} \E_B^{\m\n} +10 \E_\vp^2 
\nn\w2
&\, -32\E_\vp \E^\m{}_\m +32 \E_{\m\n} \E^{\m\n} -16 \E_\vp \E^\m{}_\m + 16 ( \E^\m{}_\m )^2 \Big] 
\end{align}
This result shows that the dualization of the Lagrangian $\cL_{BdR}$ is indeed equal to $\cL_{\alpha,\gamma}$ with $\gamma=\alpha$, and modulo field redefinitions. 

As far as supersymmetry is concerned, it also shows that the only boson that needs to be redefined is the two-form potential $B$, since terms proportional to field equations can always be set to zero in the on-shell supersymmetry transformations. Using the results for the supertransformations discussed in subsection 3.1, setting $\alpha=\gamma$, passing over to the string frame and performing the field redefinition just discussed gives the known supertransformations that leave the BdR action in $6D$ invariant; see eq. (6.3) in \cite{Chang:2021tsj}. Passing over to the string frame, in addition to \eq{mrs} also requires the relations,
\begin{align}
\psi_\mu \rightarrow \psi^\prime_\mu  &= e^{\varphi/2} ( \psi_\mu + \gamma_\mu \chi)\ , \qquad \chi \rightarrow \chi^\prime = e^{-\varphi/2} \chi\ , \qquad \epsilon \rightarrow \epsilon^\prime = e^{\varphi/2} \epsilon\ , 
\end{align}
and the resulting supertransformations are
\begin{align}
\delta e_\m{}^a
&= {\bar\e} \gamma^a \psi_\m\ , 
\nn\w2
\delta\psi_\m
&= D_\m(\omega) \epsilon  +\tfrac14 \cG_{\m\n\rh}\gamma^{\n\rh} \epsilon \ , 
\nn\w2
\delta C_{\m\n}
&= - {\bar\e} \gamma_{[\m} \psi_{\nu]} - 2 \alpha \tr\,\big( \omega_{-[\mu}\delta {\hat\omega}_{-\nu]}\big) \ , 
\nn\w2
\delta \chi &= \tfrac12 \gamma^\m \epsilon \partial_\m \vp -\tfrac{1}{12} \cG_{\m\n\rh} \gamma^{\m\n\rh} \epsilon\ ,
\nn\w2
\delta \vp &= {\bar\e}\chi\ ,
\label{6dBdRsusy}
\end{align}
where $\cG_{\m\n\rh}= 3\partial_{[\m} C_{\n\rh]} -6 \alpha\, \omega^L_{\m\n\rh}(\omega_-)$, with $\omega_- =\omega -G$ and $G=dC$.

We conclude this section by addressing whether the Lagrangian \eq{genDL} has any implication for the possibility of constructing two independent four derivative extension of heterotic supergravity in $10D$. In the case of $\gamma=\alpha$, the resulting BdR action is straightforwardly related to the ordinary dimensional reduction of the BdR action in $10D$ on $T^4$ followed by consistent truncation \cite{Chang:2021tsj}. However, if $\gamma \ne \alpha$, there seems to be an obstacle since the terms in the last line of  \eq{genDL}, to wit,
\begin{align}
\cL_{\al,\gamma}^{\rm dual}: \qquad  -\tfrac16 (\alpha-\gamma) e^{2\vp}  \tG^{\m\n\rh} \Big(G^2_{\m\alpha} G_{\n\rh}{}^\alpha -3 G_{\m\n}{}^\alpha \big( R_{\rh\alpha}-4\vp_{\rh\alpha} +4\vp_\rh\vp_\alpha\big)\Big)\ , 
\end{align}
do not seem to lift to $10D$. This provides a supportive evidence for the uniqueness of the BdR action in $10D$ as the four-derivative supersymmetric extension of heterotic supergravity in $10D$.

%%%%%%%%%%%%%%%%%%%%%%%%%%%%%%%%%%%%%%%%%%%%%%%%%%%%%%%%%%%%%%
\section{Dualization of the BdR Lagrangian in $10D$ }
%%%%%%%%%%%%%%%%%%%%%%%%%%%%%%%%%%%%%%%%%%%%%%%%%%%%%%%%%%%%%%

In this section we shall start from the BdR Lagrangian in $10D$, including fermions, and perform Hodge-dualization. This has already been achieved in \cite{BdR90}. Our aim here is to have a closer look at this dualization in the brane frame, and explore the possibility of extracting two distinct invariants in analogy with the ${\rm Riem}^2$ and Gauss-Bonnet invariants that exist in $6D$. We shall also compare the result of dualization with results obtained in superspace in \cite{Saulina:1996vn}.

The BdR Lagrangian in $10D$ has the same form as in $6D$. Up to quartic fermions, and order $\alpha'$, the BdR Lagrangian in $10D$ is given by
\be
\boxed{\begin{aligned}
\cL_{BdR}^{10D} =&  ee^{2\vp} \Bigg[ 
\tfrac{1}{4} R(\omega) + g^{\m\n} \partial_\m \vp \partial_\n \vp
- \tfrac{1}{12} \cG_{\m\n\rh} \cG^{\m\n\rh} 
\w2
& \quad  - \tfrac{1}{2} \bpsi_\m \gamma^{\m\n\rh} D_\n(\omega) \psi_\rho 
+ 2 {\bar\chi}\gamma^{\m\n}  D_\m(\omega) \psi_\n
+ 2 {\bar\chi} \gamma^\m D_\m(\omega)\chi
\w2
& \quad  - \tfrac{1}{24} \cG_{\m\n\rh} \cO^{\m\n\rh} 
- \partial_\m \vp\Bigl(  \bpsi^\m \gamma^\n \psi_\n + 2\bpsi_\n \gamma^\m \gamma^\n \chi\Bigr)
\w2
& \quad + \alpha'  \Big( - \tfrac14 R_{\m\n ab}(\omega_-) R^{\m\n ab}(\omega_-) - R_{\m\n ab} (\omega_-) D^\m (\omega_-) \big( \bpsi^a \gamma^\n \psi^b \big)
\w2
& \quad -\bpsi^{ab}\gamma^\m D_\m(\omega,\omega_-)\psi_{ab}  +\tfrac12 R_{\m\n}{}^{ab}(\omega_-)  
\bpsi_{ab} \left(\gamma^\rh \gamma^{\m\n} \psi_\rh + 2\gamma^{\m\n}\chi \right)
- \tfrac{1}{12} G_{\m\n\rh} \bpsi^{ab} \gamma^{\m\n\rh} \psi_{ab}  \Big)\,\Bigg]
\label{BdR6}
\end{aligned}}
\ee
where 
\begin{subequations}
\begin{align}
\omega_{\pm \m ab} =& \omega_{\m ab} \pm G_{\m ab}\ , \qquad G_{\m\n\rh} = 3\partial_{[\m} C_{\n\rh]}\ ,
\label{d1}\w2
\cG_{\m\n\rh} =& G_{\m\n\rh}  -6\alpha'\, \omega^L_{\m\n\rh}({\hat\omega}_-(G))\ ,
\label{d2}\w2
\omega^L_{\m\n\rh}({\hat\omega}_-) =& \tr \left(
{\hat\omega}_{-[\m} \partial_\n {\hat\omega}_{-\rh]} + \tfrac23 {\hat\omega}_{-[\m}{\hat\omega}_{-\n} {\hat\omega}_{-\rh]} \right)\ ,
\nn\w2
=& \omega^L_{\m\n\rh}(\omega_-) + \Big[ \partial_\m \big( \omega_{-\n}{}^{ab} \bpsi_a \gamma_\rh \psi_b \big) -R_{\m\n}{}^{ab}(\omega_-) \bpsi_a \gamma_\rh \psi_b \big)\Big]_{[\m\n\rh]} +\cO(\al')\ ,
\label{d3}\w2
\psi_{ab} 
=& 2 e_a{}^\m e_b{}^\n
D_{[\mu}(\omega_+)\psi_{\n]}\ , 
\label{d4}\w2
D_\m (\omega,\omega_-)\psi_{ab}
=& \left( \partial_\m 
+ \tfrac14 \omega_{\m pq} \gamma^{pq} \right) \psi_{ab} 
+\omega_{-\m a}{}^c \psi_{cb} + \omega_{-\m b}{}^c\psi_{ac}\ ,
\label{d5}\w2
\cO^{\m\n\rh} =& \Big( 
\bpsi^\s \gamma_{[\s} \gamma^{\m\n\rh} \gamma_{\tau]}\psi^\tau
+4 \bpsi_\s \gamma^{\s\m\n\rh}\chi
- 4 {\bar\chi} \gamma^{\m\n\rh} \chi \Big)\ .
\label{d6}
\end{align}
\label{dn}
\end{subequations}
It is understood that the term proportional to $\al'^2$ coming from $\cG^2$ is to be dropped, since we are considering the Lagrangian to first order in $\alpha'$. Various supercovariantizations are given by
\begin{align}
& {\hat\omega}_{\pm \m ab} = {\hat\omega}_{\m ab}\pm {\hat G}_{\m ab}\ ,
\label{oh1}\w2
& \quad
{\hat\omega}_{\m ab} = \omega_{\m ab} + \bpsi_\m \gamma_{[a}\psi_{b]} +\tfrac12\bpsi_a \gamma_\m \psi_b \ ,
\quad
{\hat G}_{\m ab} = G_{\m ab} + \tfrac32\bpsi_{[\m}\gamma_a\psi_{b]} \ .
\label{oh2}\w2
& {\hat\omega}_{-\m ab} = \omega_{-\m ab} +\bpsi_a\gamma_\m\psi_b\ ,\qquad 
{\hat\omega}_{+\m ab} = \omega_{+\m ab} +2\bpsi_\m\gamma_{[a}\psi_{b]}\ .
\end{align}
It is understood that only $\cO(\alpha')$ terms are to be kept, and that the quartic fermion terms are to be dropped in the Lagrangian \eq{BdR6}.
 
The action of the Lagrangian \eq{BdR6} is invariant under the following supersymmetry transformation rules up to ${\cal O}(\alpha')$ and cubic fermion terms,
\begin{align}
\delta e_\m{}^r
=& {\bar\e} \gamma^r \psi_\m\ , 
\nn\w2
\delta\psi_\m
=& D_\m(\omega_+(\cG)) \epsilon \ , 
\nn\w2
\delta C_{\m\n}
=& - {\bar\e} \gamma_{[\m} \psi_{\nu]} + 2 \alpha' \,\big( \omega_{-[\mu}{}^{rs}\delta {\hat\omega}_{-\nu] rs}\big) \ , 
\nn\w2
\delta \chi =& \tfrac12 \gamma^\m \epsilon \partial_\m \vp -\tfrac{1}{12} \cG_{\m\n\rh} \gamma^{\m\n\rh} \epsilon\ ,
\nn\w2
\delta \vp =& {\bar\e}\chi\ .
\label{6dsuper}
\end{align}
Next, we study the dualization of the above Lagrangian by introducing the Lagrange multiplier term  
\begin{align}
\Delta \cL^{10D} (B,C) =& \tfrac{1}{6\times 7!}  \e^{\m\n\rh\s_1...\s_7} H_{\s_1...\s_7} G_{\m\n\rh}
= \tfrac16 e\, \tH^{\m\n\rh} \big( \cG_{\m\n\rh} + 6\alpha' \omega^L_{\m\n\rh} ({\hat\omega}_-) \big)
\end{align}
where 
\be
H_{\m_1...\m_7} = 7\partial_{[\m_1} B_{\m_2...\m_7]}\ ,\qquad \tH^{\m\n\rh} = \tfrac{1}{7!} \ve^{\m\n\rh\s_1...\s_7} H_{\s_1...\s_7}\ .
\label{defs10d}
\ee

Next, we integrate over $\cG_{\m\n\rh}$. Thus we need its field equation that follows from
\begin{align}
& \cL_{BdR}^{10D} +\Delta\cL^{10D}(B,C) = \cL_{01}+ \alpha' \cL_1\ ,
\label{L1}\w4
& \cL_{01} =  ee^{2\vp} \Big[ 
\tfrac{1}{4} R(\omega) + g^{\m\n} \partial_\m \vp \partial_\n \vp
- \tfrac{1}{12} \cG_{\m\n\rh}\big( \cG^{\m\n\rh} -2e^{-2\vp} \tH^{\m\n\rh} \big)
\nn\w2
& \qquad\qquad  - \tfrac{1}{2} \bpsi_\m \gamma^{\m\n\rh} D_\n(\omega) \psi_\rho 
+ 2 {\bar\chi}\gamma^{\m\n}  D_\m(\omega) \psi_\n
+ 2 {\bar\chi} \gamma^\m D_\m(\omega)\chi 
\nn\w2
&\qquad\qquad - \partial_\m \vp\Bigl(  \bpsi^\m \gamma^\n \psi_\n + 2\bpsi_\n \gamma^\m \gamma^\n \chi\Bigr)
 - \tfrac{1}{24} \cG_{\m\n\rh} \cO^{\m\n\rh} \,\Big]
 \label{L2}\w2
& \cL_1 =  ee^{2\vp} \Big[ - \tfrac14 R_{\m\n ab}(\omega_-) R^{\m\n ab}(\omega_-)
- R_{\m\n ab} (\omega_-) D^\m (\omega_-) \big( \bpsi^a \gamma^\n \psi^b \big) 
\nn\w2
&\quad\qquad  - \bpsi^{ab}\gamma^\m D_\m(\omega,\omega_-)\psi_{ab}   +\tfrac12 R_{\m\n}{}^{ab}(\omega_-)  
\bpsi_{ab} \left(\gamma^\rh \gamma^{\m\n} \psi_\rh + 2\gamma^{\m\n}\chi \right)
\nn\w2
& \quad\qquad - \tfrac{1}{12} G_{\m\n\rh} \bpsi^{ab} \gamma^{\m\n\rh} \psi_{ab}\Big]  + e \tH^{\m\n\rh} \Big( \om^L_{\m\n\rh}({\om}_-) -R_{\m\n ab}(\omega_-) \bpsi^a \gamma_\rh \psi^b \Big) \ .
\label{L3}
\end{align}
We have collected the $\cO(\al')$ terms in which the dependence  on $\cG$ arises through the torsionful connection $\omega_-$. We are treating $\cG$ as independent variable, while $H=dB$. Thus, the field equation for $B$ gives the relation $d\cG = -\alpha' \tr (R \wedge R)$, which can be solved to give \eq{d2}. Recalling that we only work up to order $\alpha'$ Lagrangian, in expressions above $\omega_- = \omega-G$. The field equation for $\cG$ at $\cO(\al')$ following from $\int d^{10}x (\cL_{01}+\alpha' \cL_1)$ is given by 
\begin{align}
& \cG_{\m\n\rh} = e^{-2\vp} \tH_{\m\n\rh} -\tfrac14 \cO_{\m\n\rh}  + 6 \alpha' e^{-2\vp} \frac{\delta \cL_1}{\delta G^{\m\n\rh}} \ .
\label{de}
\end{align}
This equation is readily solved for $\cG$ in terms of $H$, again at $\cO(\alpha')$, as\footnote{Note that the last term produces quartic fermion terms as well but such terms are understood to be omitted throughout the paper because all results are up quartic fermion terms.}
\be
\boxed{\begin{aligned}
& \cG_{\m\n\rh} = e^{-2\vp} \tH_{\m\n\rh} -\tfrac14 \cO_{\m\n\rh}  + 6 \alpha' e^{-2\vp} \frac{\delta \cL_1}{\delta G^{\m\n\rh}}\Big|_{G=e^{-2\vp} \tH +\cO} \ .
\label{de3}
\end{aligned}}
\ee
The $G$-dependence of $\cL_1$ arises through dependence on $\omega_-(G)$ and $\psi_{ab}$ in all the terms except the term $ - \tfrac{1}{12}\al' e e^{2\vp} G_{\m\n\rh} \bpsi^{ab} \gamma^{\m\n\rh} \psi_{ab}$ where it also appears explicitly. In \cite{BdR90} it has been shown that  $\delta \cL_1 / \delta \omega_{-\m ab}$ and $\delta\cL_1 / \delta \psi_{ab}$ are proportional to field equations. Therefore, the last term \eq{de3} will be of the form $\bpsi^{ab}\gamma_{\m\n\rh} \psi_{ab} + \mbox{EOM terms}$, the details of which can be found in \cite{BdR90}. 
The supertransformations of the dualized theory are obtained by substituting for $\cG$ in the case of the fermions. As for the six-form potential $B_{\n_1...\n_6}$, its supersymmetry variation is determined by treating $\cG$ as independent field in the supersymmetry variations of $\cL_{01}+ \al' \cL_1$ and demanding that all terms proportional to $d\cG$ cancel. Up to quartic fermions, such terms are given by \cite{BdR90} 
% \hcnote{An overall minus sign was missing, and the coefficient of the last term is corrected.}
%
\begin{align}
\delta \cL_{01}\Big|_{d\cG} + \alpha' \delta\cL_1\Big|_{d\cG} =&  -  \tfrac{1}{6\times 6!} \e^{\m\n\rh\s\n_1...\n_6} \partial_\s \cG_{\m\n\rh} \Big[ \delta B_{\n_1...\n_6} 
\nn\w2
&- e^{2\vp}\Big({ 3 {\bar\e} \gamma_{[\n_1...\n_5} \psi_{\n_6]} +  {\bar\e} \gamma_{\n_1...\n_6}\chi} \Big)  - \tfrac12 \al' \frac{\delta\cL_1}{\delta\psi_{ab}}\Big|_{G=e^{-2\vp} \tH}~ \gamma_{ab\n_1...\n_6}\e\Big]\ ,
\end{align}
where \eq{dpsi} with $H$ replaced by $\cG$ has been used, and the notation $X\big|_{d\cG}$ refers to terms proportional to $d\cG$ in $X$.  Thus, in the dualized theory, and in the string frame, supertransformations up to cubic fermions and at $\cO(\al')$ are given by
\begin{align}
\delta e_\m{}^a
=& {\bar\e} \gamma^a \psi_\m\ , 
\nn\w2
\delta\psi_\m
=& D_\m(\omega)\e +\tfrac14 e^{-2\vp} \Big( \tH_{\m\n\rh} + 6 \alpha' \frac{\delta \cL_1}{\delta G^{\m\n\rh}}\Big|_{G=e^{-2\vp} \tH} \Big) \gamma^{\n\rh} \e \ , 
\nn\w2
\delta B_{\m_1...\m_6}
=& e^{2\vp}\Big({ 3 {\bar\e} \gamma_{[\m_1...\m_5} \psi_{\m_6]} +  {\bar\e} \gamma_{\m_1...\m_6}\chi} \Big) +\tfrac12 \al' \frac{\delta\cL_1}{\delta\psi_{ab}}\Big|_{G=e^{-2\vp} \tH} \gamma_{ab\n_1...\n_6}\e \ , 
\nn\w2
\delta \chi =& \tfrac12 \gamma^\m \epsilon \partial_\m \vp -\tfrac{1}{12} e^{-2\vp} \Big( \tH_{\m\n\rh} 
+6\alpha' \frac{\delta \cL_1}{\delta G^{\m\n\rh}}\Big|_{G=e^{-2\vp} \tH} \Big) \gamma^{\m\n\rh} \e\ ,
\nn\w2
\delta \vp =& {\bar\e}\chi\ .
\label{10dsuper}
\end{align}
The $\alpha'$ dependent terms turn out to be proportional to a combination of field equations which can be read of from eq. (33) of \cite{BdR90}. Substituting for $\cG$ given in \eq{de} back into the Lagrangian $\cL$ gives 
\begin{align}
\cL_{BdR}^{10D,\rm dual} =& ee^{2\vp} \biggl[ 
\tfrac{1}{4} R(\omega) + g^{\m\n} \partial_\m \vp \partial_\n \vp
 - \tfrac{1}{2\times 7!} e^{-4\vp} H_{\m_1...\mu_7} H^{\m_1...\mu_7} 
\nn\w2
& \quad  - \tfrac{1}{2} \bpsi_\m \gamma^{\m\n\rh} D_\n(\omega) \psi_\rho 
+ 2 {\bar\chi}\gamma^{\m\n}  D_\m(\omega) \psi_\n
+ 2 {\bar\chi} \gamma^\m D_\m(\omega)\chi \nonu
& \quad  - \tfrac{1}{24} e^{-2\vp}\tH_{\m\n\rh} \cO^{\m\n\rh} 
- \partial_\m \vp\Bigl(  \bpsi^\m \gamma^\n \psi_\n + 2\bpsi_\n \gamma^\m \gamma^\n \chi\Bigr)  \biggr] 
\nn\w2
& \quad -\tfrac{1}{4\times 6!} \alpha'\, \ve^{\m\n\rh\s\al_1...\al_6} B_{\al_1...\al_6} R_{\m\n}{}^{ab} ({\hat\omega}_-) R_{\rh\s ab} ({\hat\omega}_-)\Big|_{\cG=e^{-2\vp} \tH + \cO } 
\nn\w2
& \quad  +\alpha' e e^{2\vp} \Bigl[ - \tfrac14 R_{\m\n ab}({\hat\omega}_-) R^{\m\n ab}({\hat\omega}_-) - \bpsi^{ab}\gamma^\m D_\m(\omega,\omega_-)\psi_{ab}
\nn\w2
& \quad +\tfrac12 R_{\m\n}{}^{ab}(\omega_-)  
\bpsi_{ab} \left(\gamma^\rh \gamma^{\m\n} \psi_\rh + 2\gamma^{\m\n}\chi \right)
- \tfrac{1}{12} \cG_{\m\n\rh} \bpsi^{ab} \gamma^{\m\n\rh} \psi_{ab}  \Bigr]_{\cG=e^{-2\vp} \tH + \cO }\ ,
\label{dualBdRv1}
\end{align}
in agreement with \cite{BdR90}. To explore further the structure of this result, using \eq{RH}, \eq{LCSH} (with $H$ replaced by $G$) and \eq{r2}, we obtain for the the bosonic part of the Lagrangian
\begin{align}
\cL_{BdR}^{10D,\rm dual} =& ee^{2\vp} \Bigg[ 
\tfrac{1}{4} R(\omega) + g^{\m\n} \partial_\m \vp \partial_\n \vp
 - \tfrac{1}{2\times 7!} e^{-4\vp} H_{\m_1...\mu_7} H^{\m_1...\mu_7} \Big]
\nn\w2
& + \alpha' \Big( - \tfrac14 R_{\m\n\rh\s} R^{\m\n\rh\s} + 6 e^{- 4 \vp} R^{\m\n} H^2_{\m\n} - 3 e^{- 4 \vp} R H^2 - \tfrac32 e^{- 4 \vp} R^{\m\n\rh\s} H_{\m\n, \rh\s} 
\nn\w2
&\, - \tfrac16 e^{- 8 \vp} H_4 - \tfrac23 e^{- 8 \vp} H^2_{\m\n} H^{2 \m\n} + \tfrac{14}{3} e^{- 8 \vp} ( H^2 )^2 
\nn\w2
&\, + \tfrac{2}{7!} e^{- 4 \vp} ( D_\m H_{\n_1 \cdots \n_7} ) ( D^\m H^{\n_1 \cdots \n_7} ) + 8 e^{- 4 \vp} H^2 \vp^2 - 4 e^{- 4 \vp} \vp^\m D_\m H^2 
\nn\w2
&\, + e^{- 2 \vp} \tH^{\m\n\rh} \omega_{\m\n\rh}^L(\omega) + \tfrac{1}{360} e^{- 6 \vp} \tH^{\m\n\rh} H_\m{}^{\la_1 \cdots \la_6} D_\n H_{\rh \la_1 \cdots \la_6} \Big) \Bigg] \ , 
\label{10DdualBS}
\end{align}
where the products of $H$'s are defined as
\begin{equation}
H_{\m\n, \rh\s} := \tfrac{1}{5!} H_{\m\n\la_1 \cdots \la_5} H_{\rh\s}{}^{\la_1 \cdots \la_5} , \quad H^2_{\m\n} := \tfrac{1}{6!}H_{\m\la_1 \cdots \la_6} H_\n{}^{\la_1 \cdots \la_6} , \quad H^2 := \tfrac17 H^2_{\m\n} g^{\m\n} \ . 
\end{equation}
We have also used the following two lemmas:
\begin{align}
\tH_{\m\n}{}^\la \tH_{\rh\s\la} =&\,\Big( - H_{\m\n, \rh\s} + 4 H^2_{\m\rh} g_{\n\s} - 2 H^2 g_{\m\rh} g_{\n\s} \Big)  \Big|_{[\m\n], [\rh\s]} 
\nn\w2
\tH_\m{}^{\rh\s} \tH_{\n\rh\s} =&\, 2 H^2_{\m\n} - 2 H^2 g_{\m\n} \ .
\end{align}
It is useful to study the ordinary dimensional reduction of the Lagrangian \eq{10DdualBS} on 4-torus, followed by a consistent truncation in which the only kept bosonic fields are $(g_{\m\n}, B_{\m\n}, \vp)$. This is done in Appendix B.

Next, we go over to the brane frame \cite{Duff:1990wv} by rescaling the metric as
\be
g_{\m\n} \to  g'_{\m\n} = e^{-2\vp/3} g_{\m\n}\ ,
\ee
thereby obtaining the Lagrangian
\begin{equation}
\boxed{
\begin{aligned}
\cL^{10D, \rm dual}_{BdR} =&\, e\, e^{- 2\vp/3} \Big[ \tfrac14 R - \tfrac{1}{2 \times 7!} H_{\m_1 \cdots \m_7} H^{\m_1 \cdots \m_7} \Big] 
\w2
&\, + \alpha'\, e\, \Big[ - \tfrac14 R^{\m\n\rh\s} R_{\m\n\rh\s} + 6 H^2_{\m\n} R^{\m\n} - 3 H^2 R - \tfrac32 R^{\m\n\rh\s} H_{\m\n, \rh\s} 
\w2
&\, - \tfrac23 H^2_{\m\n} H^{2 \m\n} - \tfrac16 H_4 + \tfrac{14}{3} ( H^2 )^2 + \tfrac{2}{7!} ( D_\m H_{\n_1 \cdots \n_7} ) D^\m H^{\n_1 \cdots \n_7} 
\w2
&\, + 6 \vp^\m D_\m H^2  + \tfrac49 H^2_{\m\n} \vp^\m \vp^\n + \tfrac{16}{3} H^2_{\m\n} \vp^{\m\n} + \tfrac{22}{9} H^2 \vp^2 
\w2
&\, - \tfrac29 R_{\m\n} \vp^\m \vp^\n - \tfrac23 R_{\m\n} \vp^{\m\n} + \tfrac19 R \vp^2 
- \tfrac{16}{27} \vp_{\m\n} \vp^\m \vp^\n 
\w2
&\, - \tfrac89 \vp_{\m\n} \vp^{\m\n} - \tfrac49 ( \vp^2 )^2 + \tfrac{16}{27} \vp^2 \vp^\m{}_\m - \tfrac19 ( \vp^\m{}_\m )^2 
\w2
&\,  + \tfrac{2}{6!} \tH^{\m\n\rh} H_\m{}^{\s_1 \cdots \s_6} D_\n H_{\rh \s_1 \cdots \s_6} - \tfrac43 \tH^{\m\n\rh} \vp^\s H_{\m\n, \rh\s} + \tH^{\m\n\rh} \omega^L_{\m\n\rh}(\omega) \Big] 
\label{BdRDual10D}
\end{aligned}}
\end{equation}
As to the supertransformations, we also need to redefine the fermions as 
\be
 \psi_\m \to  \psi_\m' = e^{-\vp/6} \big( \psi_\m -\tfrac13 \gamma_\m \chi \big)\ ,\qquad \chi \to  \chi'= e^{\vp/6} \chi\ ,\qquad \epsilon \to \epsilon' = e^{-\vp/6} \epsilon\ .
\ee
Dropping a local Lorentz transformation of $e_\m{}^a$, the supertransformations \eq{10dsuper} expressed in the brane frame, and  up to cubic fermions, take the form \cite{BdR90}
\begin{align}
\delta e_\m{}^a
=& {\bar\e} \gamma^a \psi_\m\ , 
\nn\w2
\delta\psi_\m
=& D_\m(\omega)\e + \tfrac{1}{72} \tH_{abc} \big( 3\gamma^{abc}\gamma_\m +\gamma_\m \gamma^{abc}\big)\e + {\rm EOMs}\ ,
\nn\w2
\delta B_{\m_1...\m_6}
=& 3 {\bar\e} \gamma_{[\m_1...\m_5} \psi_{\m_6]} +{\rm EOMs}\ , 
\nn\w2
\delta \chi =& \tfrac12 \gamma^\m \epsilon \partial_\m \vp -\tfrac{1}{12} \tH_{\m\n\rh} \gamma^{\m\n\rh} \e +{\rm EOMs}\ ,
\nn\w2
\delta \vp =& {\bar\e}\chi\ .
\label{10dsuper2}
\end{align}
%

%%%%%%%%%%%%%%%%%%%%%%%%%%%%%%%%%%%%%%%%%%%%%%%%%%%%%%%
\section{Comparison with results in superspace}
%%%%%%%%%%%%%%%%%%%%%%%%%%%%%%%%%%%%%%%%%%%%%%%%%%%%%%%
%
The dualization of heterotic supergravity in $10D$ has also been studied in superspace. It is useful to compare the results described above with those obtained from superspace considerations. Starting from the two-form formulation, the key equations for the superspace description are the Bianchi identities 
\be
DT^A= R^A{}_B \wedge E^B\ ,\qquad D\cG = \alpha' \tr (R\wedge R)\ .
\label{SG}
\ee
With a particular set of constraints these were solved in \cite{Bonora:1986ix, Bonora:1987xn, DAuria:1987tdr, Raciti:1989je, Bonora:1990mt, Bonora:1992tx,Fre:1991ef, Pesando:1992pa} \footnote{These BI's have been analyzed in superspace also for $N=(1,0), 6D$ supergravity in \cite{DallAgata:1997yqq}.}, where the consistency of the BI's was proven to all orders in $\alpha'$. In this approach the dimension zero torsion component is taken to be $T_{\al\beta}^a= \gamma^a_{\al\beta}$ but certain other components are deformed by $\al'$ dependent terms. In particular the following relation (in our notation) arises
\be
\cG_{abc} = e^{-2\vp} T_{abc} + \al' W_{abc} (T)\ ,
\label{GBI}
\ee
where $W_{abc}$ is a nonlinear function of the torsion superfield $T_{abc}$ which can be found in the papers referred to above. To obtain the deformed equations of motion, one solves for $T_{abc}$ in terms of $G_{abc}$ order by order in $\alpha'$, and uses the result in the supertorsion BI's. The resulting equations of motion were obtained at $\cO(\al')$ in\cite{Fre:1991ef, Pesando:1992pa}. These equations apparently have not been compared with those which arise from the BdR action. While they are expected to agree at $\cO(\alpha')$, it is an open question whether equivalence holds to all orders in $\alpha'$. This approach has been updated in \cite{Lechner:2008uz} where relationship to another approach by \cite{Bellucci:1988ff, Bellucci:1990fa} which focuses on order by order in $\al'$ analysis (without addressing fully the question of the consistency of the entire procedure) was clarified. Interestingly, the formulation of \cite{Lechner:2008uz} is such that the Gauss-Bonnet action appears as part of the bosonic action. The full four-derivative action in this framework has not been worked out but it is expected to be related to the result that follow from \cite{Bonora:1986ix, Bonora:1987xn, DAuria:1987tdr, Raciti:1989je, Bonora:1990mt, Bonora:1992tx} by field redefinitions. 

The adopted constraints on torsion and $\cG$ in proving the consistency of \eq{SG} can yield the deformed equations of motion to any order in $\al'$. However, this framework does not capture the most general supersymmetric deformation. For example, at order $\cO(\al'^3)$, deformations involving $(R^2)^2$ but not $R^4$ will arise. To get the latter, one can either deform the constraint on $T^a_{\al\beta}$ to include a tensor in $1050$ dimensional representation of $SO(9,1)$ \cite{Bellucci:2006cx,OReilly:2006eeg,Nilsson:1986cz,Candiello:1994ew,Howe:2008vb,Lechner:2010ti} or take $\cG_{\alpha\beta\gamma}$ to be nonvanishing\cite{Lechner:1987ip}.

Putting aside the question of dualization, heterotic supergravity directly in the six-form formulation in superspace including $\alpha'$ corrections was studied in \cite{Gates:1985wh, Gates:1986is, Gates:1986tj, Nishino:1986mj, Nishino:1990ky} where partial results were obtained. A more complete  treatment which builds especially on the results of \cite{Nishino:1990ky} appeared in \cite{Terentev:1993wm, Terentev:1994br, Zyablyuk:1994xk, Saulina:1995eq, Saulina:1996vn}, where the dualization phenomenon in superspace, suggested in \cite{DAuria:1987qjh}, was spelled out as well. Here we shall focus on the key results of \cite{Saulina:1996vn} where the equations of motion deduced from superspace were also integrated into an action for the bosonic fields, and we shall compare the result with ours.

The super BI's for supertorsion  $T_{MN}^A$ and the super seven-form $H_7=DB_6$ are given by
\begin{align}
DT^A &= R^A{}_B \wedge E^B\ , \qquad  DH_7 =0\ .
\label{DBI}
\end{align}
Note that the BI for $H_7$ does not acquire $\alpha'$ deformation, unlike the BI for $\cG$ in \eq{GBI}. The BI's \eq{DBI} are solved by (see \cite{Saulina:1996vn} and references therein)
\begin{align} 
& T_{\alpha\beta}{}^c = \gamma^c_{\alpha\beta}\ ,\qquad  T_{a\beta}{}^\gamma = \tfrac{1}{(72)^2} T_{bcd} \left(\gamma^{bcd}\gamma^a \right)_\beta{}^\gamma\ , \qquad  T_{\alpha b}{}^c =0\ ,\qquad  T_{\alpha\beta}{}^\gamma=0\ ,
\nn\w2
& H_{a_1...a_5 \alpha\beta} = -\left(\gamma_{a_1...a_5}\right)_{\alpha\beta}\ ,\qquad \boxed{H_{a_1...a_7}= \tfrac{1}{6!} \e_{a_1...a_7}{}^{abc} T_{abc}} \ , 
\nn\w2
& \mbox{other components of}\ \  H_7=0\ ,
\label{ssc}
\end{align}
together with a scalar superfield $\phi$ with
\be
D_\alpha \phi = \chi_\alpha\ ,\qquad D_\alpha \chi_\beta  = \tfrac12 \gamma^a_{\alpha\beta} D_a \phi +\left(-\tfrac{1}{36} \phi T_{abc} +\alpha' A_{abc} \right) \left(\gamma^{abc}\right)_{\alpha\beta}\ ,
\label{dchi}
\ee
where $D_a$ is covariant derivative with bosonic torsion, and $A_{abc}$ is a crucial superfield which governs the $\alpha'$ deformation given by~\cite{Saulina:1996vn}\footnote{Certain terms for $A_{abc}$ and their implications for the $\alpha'$ corrections were considered in \cite{Gates:1985wh, Gates:1986is, Gates:1986tj, Nishino:1986mj, Nishino:1990ky}.} 
\begin{align}
A_{abc}& = \Big[-\tfrac{1}{18} \Box T_{abc} +\tfrac{1}{36} D^d T_{da,bc} -\tfrac{1}{36} T^{de}{}_a D_b T_{c de}
-\tfrac{5}{1944} T^2 T_{abc}
\nn\w2
& -\tfrac{5}{108} T^2_{da} T_{bc}{}^d +\tfrac{5}{54} T^3_{abc} -\tfrac{1}{3888} \e_{abc}{}^{a_1...a_7} T_{a_1...a_3} D_{a_4} T_{a_5...a_7} 
\nn\w2
& -\tfrac{1}{48} T_{a_1 a_2}{}^\alpha \left( 2\gamma_{abc} \eta^{a_1b_1} \eta^{a_2 b_2} + \gamma^{a_1}\gamma_{abc} \gamma^{b_1} \eta^{a_2 b_2} +24 \gamma^{a_1} \gamma_c \gamma^{b_1} \delta_a^{a_2} \delta_b^{b_2} \right) T_{b_1 b_2}{}^\beta \Big]_{[abc]}\ ,
\end{align}
where $T_{abc}= T_{[abc]}$ and $T_{ab}{}^\alpha$ is the gravitino curvature, and 
\be 
T_{ab,cd} := T_{ab}{}^e T_{cde}\ ,\ \ T^2_{ab}:= T_a{}^{cd} T_{bcd}\ ,\ \  T^3_{abc} := T_{a d_1d_2} T_b{}^{d_2d_3} T_{cd_3}{}^{d_1}\ ,\ \  T^2 := T_{abc} T^{abc}\ .
\ee
It is noteworthy that the solution is an exact one, even though there is an $\alpha'$ dependent deformation. The EOM's that result from the analysis of the superspace BI's are also given in \cite{Saulina:1996vn} in terms of superfields whose lowest order components in $\theta$ expansion are the supergravity multiplet of fields.  For a more detailed explanation of how the EOMs are obtained in superspace, see \cite{Terentev:1994br}. These equations imply an action with $\alpha' Riem^2$ term, and yet their supersymmetry is realized exactly. No higher than first order in $\alpha'$ terms arise in supersymmetric variations of these EOM's since, as can be seen in \cite{Saulina:1996vn}, the $\alpha'$ dependent terms do not involve the dilatino $\chi$ which is the only field that develops $\alpha'$ deformation; see \eq{dchi}.

A bosonic Lagrangian which yields these EOM's can only be determined up to squares of the lowest order (i.e. two-derivative) EOM's. With this understood, the resulting bosonic Lagrangian is found to be \cite{Saulina:1996vn}
\footnote{In converting the conventions of \cite{Saulina:1995eq} ours, we first let $\omega_\m{}^{ab} \to - \omega_\m{}^{ab}$, and then let $\eta_{ab} \to -\eta_{ab}, \e_{a_1...a_{10}} \to -\e_{a_1...a_{10}},  M_{\m\n\rh} \to 2 \tH_{\m\n\rh}$, $\tilde{\phi} \to e^{- \tfrac23 \vp}$, $k_g \to \alpha'$, and $\cL \to 4 \cL$. Note also that the term $\tfrac{1}{162} ( M^2 )^2$ term in (4.10) of \cite{Saulina:1995eq} should be absent, as noted later in\cite{Saulina:1996vn} as well.}
\begin{align}
\cL_{STZ} &= e e^{-2\vp/3} \big[ \tfrac14 R +\tfrac{1}{12} \tH^{\m\n\rh} \tH_{\m\n\rh}\big]
\nn\w2
& + e \alpha'\big[ - \tfrac14 R^{\m\n\rh\s} R_{\m\n\rh\s} + \tfrac12 R^{\m\n} R_{\m\n} - \tfrac{1}{4 \times 6!} \ve^{\m\n\rh\s\la_1 \cdots \la_6} R_{\m\n}{}^{ab} R_{\rh\s ab} B_{\la_1 \cdots \la_6} 
\nn\w2
& + \tfrac12 R^{\m\n} ( \tH^2 )_{\m\n} - \tfrac16 \tH^{\m\n\rh} D^\s D_\s \tH_{\m\n\rh} +  ( D^\s \tH^{\m\n\rh} ) \tH_{\m\n, \rh\s} - \tfrac16 \tH^{\m\n, \rh\s} \tH_{\m\rh, \n\s} \big]
\ .
\end{align}
For completeness, we also express this Lagrangian in terms of the seven-form field strength,
\begin{align}
\cL_{STZ} =&\, e e^{- \tfrac23 \vp} \Big[ \tfrac14 R - \tfrac{1}{2 \times 7!} H_{\m_1 \cdots \m_7} H^{\m_1 \cdots \m_7} \Big] 
\nn\w2
&\, + \tfrac14 e \alpha' \Big[ - R^{\m\n\rh\s} R_{\m\n\rh\s} + 2 R^{\m\n} R_{\m\n} + 4 R^{\m\n} H^2_{\m\n} - 4 R H^2 
\nn\w2
&\, - \tfrac{4}{7!} ( D_\m H_{\n_1 \cdots \n_7} ) D^\m H^{\n_1 \cdots \n_7} + \tfrac{16}{3} H^2_{\m\n} H^{2 \m\n} - \tfrac23 H_4 - \tfrac{40}{3} ( H^2 )^2 
\nn\w2
&\, + 4 \tH^{\m\n\rh} D^\s H_{\m\n, \rh\s} + 4 \tH^{\m\n\rh} \omega^L_{\m\n\rh}(\omega) \Big] \ . 
\end{align}
The supertransformation  resulting from the constraints \eq{ssc} are \cite{Zyablyuk:1994xk} (up to cubic fermions here)
\begin{align}
\delta e_\m{}^a =& {\bar\e} \gamma^a \psi_\m\ ,
\nn\w2
\delta \psi_\m =& D_\m \e -\tfrac{1}{72} T_{abc} \big( 3 \gamma^{abc}\gamma_a +\gamma_a \gamma^{abc} \big) \e\ ,
\nn\w2
\delta B_{\m_1...\m_6} =& 3 {\bar\e} \gamma_{[\m_1...\m_5} \psi_{\m_6]}\ ,
\nn\w2
\delta\chi =& \tfrac12 \gamma^\m \e \partial_\m \phi + \left( -\tfrac{1}{36} \phi T_{abc} + \alpha' A_{abc}\right) \gamma^{abc} \e\ ,
\nn\w2
\delta \phi =& {\bar\e} \chi\ ,
\label{STZsusy}
\end{align}
where it is understood that $\phi \to e^{-2\vp/3}$ and $\chi \to e^{-2\vp/3} \chi$. These are also understood to be valid up to the lowest order EOMs. It has been shown in \cite{Saulina:1996vn} that the algebra closes on-shell, and that the closure functions are $\alpha'$ independent. Thus, the closure of the algebra is not a statement up to order $\alpha'$ but an exactly valid statement. The fact that $A_{abc}$ obeys the relation  $DA_{abc}= \gamma_{abc}{}^{de} X_{de}$ where $X_{de}{}^\alpha$ is an arbitrary function \cite{Saulina:1996vn} is behind this property.

Comparing the Lagrangian $\cL_{STZ}$ with the bosonic sector of the dual of the BdR Lagrangian in $10D$ \eq{BdRDual10D}, which was obtained by solving the duality equation to order $\alpha'$, we see that they differ by many terms. To determine the nature of these terms, we consider the lowest order field equations
\begin{equation}
\E_{\m\n} \equiv e^{-1} e^{\tfrac23 \vp} \frac{\del \cL}{\del g^{\m\n}} \ , 
\qquad 
\E_\vp \equiv e^{-1} e^{\tfrac23 \vp} \frac{\del \cL}{\del \vp} \ , 
\qquad 
\E_B^{\m_1 \cdots \m_6} \equiv 6! e^{-1} \frac{\del \cL}{\del B_{\m_1 \cdots \m_6}}\ ,
\end{equation}
and compute the analogs of the relations \eq{lemmas} for this case. In particular, we have
\begin{align}
\tfrac{1}{7!} ( D_\m H_{\n_1 \cdots \n_7} ) D^\m H^{\n_1 \cdots \n_7} =&\, \tfrac{1}{2} R^{\m\n\rh\s} H_{\m\n, \rh\s} - 2 H^2_{\m\n} H^{2 \m\n} + \tfrac{26}{3} ( H^2 )^2 - \tfrac49 H^2_{\m\n} \vp^\m \vp^\n + \tfrac29 H^2 \vp^2 
\nn\w2
&\, - 4 H^2_{\m\n} \E^{\m\n} + \tfrac{26}{9} H^2 \E^\m{}_\m + \tfrac{11}{3} H^2 \E_\vp + \tfrac{2}{3 \times 6!} e^{\tfrac23 \vp} \vp^\m H_{\m \n_1 \cdots \n_6} \E_B^{\n_1 \cdots \n_6} 
\nn\w2
&\, + \tfrac{1}{6!} e^{\tfrac43 \vp} \E^B_{\m_1 \cdots \m_6} \E_B^{\m_1 \cdots \m_6} \ .
\label{newids}
\end{align}
Using these identities we find that the difference between the bosonic part of the Lagrangians $\cL^{10D, \rm dual}_{BdR}$ and $\cL_{STZ}$ is given by
\begin{align}
&\cL^{10D, \rm dual}_{BdR} -\cL_{STZ} =\, e\, \al' \Big[  4 H^2 \E_\vp -\tfrac23 \vp^2 \E_\vp - 8 \E_{\m\n} \E^{\m\n}
\nn\w2
&\,  + \tfrac49 ( \E^\m{}_\m )^2 + 4\E^\m{}_\m \E_\vp - 13 \E_\vp^2 
+ \tfrac{3}{6!} e^{\tfrac43 \vp} \E^B_{\m_1 \cdots \m_6} \E_B^{\m_1 \cdots \m_6} \Big] \ .
\label{dif}
\end{align}
Thus we are left with terms that vanish on-shell, which imply that the two actions are related to each other by field redefinitions. Since the Lagrangian $\cL_{STZ}$ is given up to terms that are squares of the EOMs, the relevant field redefinition to consider in comparing it with $\cL_{BdR}$ is 
\begin{align}
 \vp & \to \vp +\alpha' e^{2\vp/3}\left(4H^2-\tfrac23 \vp^2\right)\ ,
 \label{vpr}
\end{align}
up to terms that are bilinear in fermions. Such terms are not available since the part of $\cL_{STZ}$ that contain the fermionic fields has not been given in \cite{Saulina:1996vn}. Consequently, the expected redefinition of the dilatino $\chi$ is not available either, we are not in a position to compute the result of the redefinitions of $\vp$ and $\chi$ in the supertransformation rules \eq{10dsuper2} to compare the result with those of $STZ$ given in \eq{STZsusy}. At any rate, the result \eq{10dsuper2} for $\delta\chi$ cannot produce the STZ result because the latter is exact in $\alpha'$ while the former is an order $\alpha'$ result. To achieve a proper comparison, the dualization of the BdR action in two-form formulation to all orders in $\alpha'$ is needed. Such a results is not available. Nonetheless, a conjectured solution may be envisaged in superspace as follows.

In superspace, leaving the solution of the BI's reviewed above intact, one can also construct a super three-form $\cG$ which obeys the super BI \eq{SG} as \cite{Saulina:1996vn}
\begin{align}
\cG_{\al\beta\gamma} =& 0\ ,
\nn\w2
\cG_{\al\beta a}=& \phi \left(\gamma_a\right)_{\al\beta} +\alpha'\,U_{\alpha\beta a}\ ,
\nn\w2
\cG_{\al bc} =& -\left(\gamma_{bc} \chi\right)_\alpha +\al'\,U_{\al bc}\ ,
\nn\w2
\cG_{abc}=& -\phi T_{abc} +\al'\, U_{abc}\ ,
\label{Geq}
\end{align}
where\cite{Saulina:1996vn}
\begin{align}
U_{abc}=& \Big[-2 \Box T_{abc} -6 D^d T_{da,bc} -6 T^{de}{}_a D_b T_{cde}
-6 {\cal R}^{de}{}_{ab} T_{cde} -6 {\cal R}_{da} T_{bc}{}^d +4T^3_{abc}
\nn\w2
& - T_{a_1 a_2}{}^\alpha \Big( \gamma_{abc} \eta^{a_1b_1} \eta^{a_2 b_2} + \gamma^{a_1}\gamma_{abc} \gamma^{b_1} \eta^{a_2 b_2} +12 \gamma^{a_1} \gamma_c \gamma^{b_1} \delta_a^{a_2} \delta_b^{b_2} 
\nn\w2
& + 12 \delta_a^{a_1} \delta_b^{b_1} \eta^{a_2 b_2} \gamma_c + 6 \delta_a^{a_1} \delta_b^{b_1} \delta_c^{a_2} \gamma^{b_2} \Big) T_{b_1 b_2}{}^\beta \Big]_{[abc]}\ ,
\end{align}
and the expressions for $U_{\al\beta a}$ and $U_{\al bc}$, which are functions of $T_{abc}$ and $T_{ab}{}^\al$, can be found in \cite{Saulina:1996vn}. The last equation in \eq{Geq} is expected to be equivalent to \eq{GBI} upon field redefinitions, and it also represents the duality relation between the two-form and six-form formulations as can be seen by  substituting $T_{abc} = \tH_{abc}$ from \eq{ssc} into this relation, which now takes the form
\be 
\cG_{abc}= -\phi \tH_{abc} +\al'\, U_{abc}\Big|_{T_{def}=\tH_{def}}\ .
\label{STZduality}
\ee
Solving for ${\widetilde H}_{abc}$ order by order in $\alpha$ and substituting the result into the EOM's obtained by STZ in \cite{Saulina:1996vn} is expected, though not proven, to generate the EOM's of BdR to all orders in $\alpha'$, just as solving for $T_{abc}$ in \eq{GBI} is expected, but not proven, to lead to the same result upon field redefinitions, as explained above. It would be interesting to reduce the duality equation \eq{STZduality} on $T_4$ and consistently truncate to $N=(1,0)$ supersymmetry, and then compare with \eq{AGDE} for $\gamma=\alpha$. The terms coming from $U_{abc}$ in such a reduction would give rise to terms similar to those arising in \eq{AGDE}. However, it must be kept in mind that while \eq{STZduality} is an exact relation, \eq{AGDE} holds up to first order in $(\alpha, \gamma)$. Therefore, the relation between the for \eq{STZduality} and \eq{AGDE} remains to be understood\footnote{One way to obtain equations of motion, and duality relation, that would be exact at order $(\alpha, \gamma)$ is to start from the off-shell Lagrangian \eq{2pL} and  eliminate only the auxiliary field $E_{\mu\nu\rho\sigma}$ discussed in Section 2 which has algebraic equation of motion~\cite{Bergshoeff:2012ax}. However, there would be  terms that depend on the auxiliary field $V_\mu^{ij}$, which do not seem to be accommodated in the dimensional reduction of the STZ results in the six-form formulation.}.

%%%%%%%%%%%%%%%%%%%%%%%
\section{Conclusions}
%%%%%%%%%%%%%%%%%%%%%%%

In this paper we have examined in detail the  four-derivative extensions of heterotic supergravities and their dualization  in six and ten dimensions.  In $6D$ the two known four-derivative invariants are off-shell supersymmetric. To dualize them, we first eliminated the auxiliary fields order by order in a derivative expansion, and then employed the time-honored Lagrange multiplier method, which avoids the far more tedious method of integrating out the dualized equations of motion into an action. We have also treated the issue of field redefinitions in a systematic manner by determining the dependence of the dualized actions on the lowest order equations of motion. Consequently the Lagrangian depends on product of fields which are differentiated at most once. 

We have shown that the two four-derivative extensions of heterotic supergravity in $6D$  cannot be related to each other by any field redefinitions. We have dualized them separately,  thereby obtaining a two parameter $(\alpha,\gamma)$  dual theory with Lagrangian $\cL^{\rm dual}_{\alpha,\gamma}$. We have shown that $\cL^{\rm dual}_{\alpha,\alpha}$ is the BdR Lagrangian upon field redefinitions that are spelled out. We have also shown that  $\cL^{\rm dual}_{\alpha,-\alpha}$ is a four-derivative extension that contains no curvature squared terms. 

We have highlighted the question of whether the BdR action in $10D$ is a unique four-derivative extension of heterotic supergravity. The existence of two such extension in $6D$ motivated us to examine more closely the dualization of BdR supergravity in $10D$ \cite{Bergshoeff:1990ax}. We have noted that while the reduction on 4-torus followed by consistent truncation to $N=(1,0)$ gives  the BdR action in $6D$, we cannot represent the dual of the BdR action in $10D$ as sum of two distinct four-derivative extensions, since certain $6D$ terms we have identified do not admit a lift to $10D$. 
Working out the details of the dualized BdR action in $10D$ has also made it possible to compare the result with that of \cite{Saulina:1996vn} obtained in superspace. We have shown that the bosonic sectors agree after suitable field redefinition that are spelled out. 
 
 While our results suggest that the BdR action may indeed be the unique four-derivative extension of heterotic supergravity in $10D$, a rigorous proof would be desirable. The general Noether procedure that does not rely on the trick of replacing the Yang-Mills field with the Lorentz connection may be forbiddingly difficult, but the superspace approach, in particular in the dual six-form formulation,  may prove to be more effective. We have reviewed various aspects of the superspace approach in the previous section. 

 The inclusion of the Yang-Mills sector both in six and ten dimensions, is straightforward at order $\alpha'$ since the two-derivative action arises at this order. More interesting and challenging problem is to construct the higher derivative couplings of heterotic supergravity to hypermultiplets and tensor multiplets in $6D$. The ordinary dimensional reduction of the $10D$ BdR action  on $T^4$, followed by consistent truncation has been shown to yield the couplings of hypers that parametrize $SO(4,4)/(SO(4)\times SO(4))$~\cite{Chang:2021tsj}. Couplings of more general quaternionic Kahler manifolds are currently under investigation \cite{CST}, and we expect that not all will be obtainable from the compactifications of string theory. This makes them a fertile and interesting arena to investigate the necessity of string theory miracles, should inconsistencies be identified in such a rich landscape of higher derivative couplings in $N=(1,0)$ supergravity in six dimensions.

\subsubsection*{Acknowledgements}

We thank Eric Bergshoeff, Daniel Butter, Michael Duff, Paul Howe, Kurt Lechner, Bengt Nilsson, Mehmet Ozkan and Yi Pang for helpful discussions. The work of ES is supported in part by the NSF grants PHY-1803875 and PHYS-2112859, and the work of HC is supported in part by the Mitchel Institute for Physics and Astronomy.  

\begin{appendix}

%%%%%%%%%%%%%%%%%%%%%%%%%%%%%%%%%%%%%%%%%%%
\section{Conventions and some identities}
%%%%%%%%%%%%%%%%%%%%%%%%%%%%%%%%%%%%%%%%%%%
In our conventions $\{\gamma_a,\gamma_b\}=2\eta_{ab}$ with $\eta_{\m\n} =(-,+,...+)$, and 
\be
\gamma^{\m_1..\m_n} \gamma_7 = -\tfrac{1}{(6-n)!} {\varepsilon}^{\m_n...\m_1 \n_1...\n_{6-n}} \gamma_{\n_1...\n_{6-n}}\ ,\qquad \gamma^{\m_1..\m_n} \gamma_{11} = \tfrac{1}{(10-n)!} {\varepsilon}^{\m_n...\m_1 \n_1...\n_{10-n}} \gamma_{\n_1...\n_{10-n}}\ .
\ee

Some useful identities that have been used in the calculations are given by
\begin{align}
& \tH^{\m\n\rh} R_{\m\n}{}^{ab} H_{\rh ab} = \tH^{\m\n\rh}  H_{\m\n}{}^\s R_{\rh\s}\ ,
\label{id1}\w2
& \tH^{\m\n\rh} H_{\m a, \n b} H_\rh{}^{ab} = \tfrac12 \tH^{\m\n\rh}  H_{\m\n}{}^\s  H^2_{\rh\s} \ ,
\label{id2}\w2
& \tH^{\m\n\rh} H_{\m\,ab} H_{\n\rh}{}^a \vp^b =0\ ,
\label{id3}\w2
& \int d^6 x\, H^{\al\beta\m} H_\al{}^{\n\rh} D_\beta \tH_{\m\n\rh} = \int d^6 x\,  \tH^{\m\n\rh} H_\m{}^{\al\beta} D_\n H_{\rh\al\beta}\ , 
\label{id4}
\end{align}
where $\tH^{\m\n\rh}= \tfrac{1}{3!}\ve^{\m\n\rh\s\la\tau} H_{\s\la\tau}$. The first one can be derived from the identically zero expression,
\be
\ve^{[\m\n\rh\s\la\ta} H_{\s\la\ta} R_{\m\n}{}^{\alpha]\beta} H_{\rh\alpha\beta} =0\ ,
\ee
and the second one from a similar expression with $R_{\m\n \al\beta} $ replaced by $H_{\m \al,\n \beta}$. The last two identities 
can be derived from 
\begin{align}
& \ve^{[\m\n\rh\s\la\ta} H_{\s\la\ta} H^{\al]\beta}{}_\m H_{\beta\n\rh}  \vp_\al=0\  ,
\nn\w2
& 
\ve^{[\m\n\rh\s\la\ta} H_{\s\la\ta} H^{\al]\beta}{}_\m D_\n H_{\rh\alpha\beta} =0\ .
\end{align}
In $10D$ we have used the lemmas
\begin{align}
& \tH^{\m\n\rh} \vp^\s H_{\m\n, \rh\s}=0\ ,
\nn\w2
&\varepsilon_{\nu\rho\sigma\tau\n_1\cdots\n_6} \tilde{H}^{\mu\nu\rho} 
\tilde{H}_\mu{}^{\sigma\tau} = -12\tH_{\m\n[\n_1} H^{\m\n}{}_{\n_2 \cdots \n_6]}\ .
\end{align}
The first identity is easy to deduce from $H_{[\m\n,\rh\s]}=-\tH_{[\m\n,\rh\s]}$.  Another useful lemmas are given by
\begin{align}
\del  \big(e e^{c \vp} R \big) =&\, e^{c\vp} \big(R_{\m\n}-\frac12 g_{\m\n} R - c^2 \vp_\m \vp_\n - c \vp_{\m\n} + c^2 \vp^2 g_{\m\n} + c g_{\m\n} \vp^\rh{}_\rh \big) \del g^{\m\n} \ ,
\nn\w2
\omega^L_{\m\n\rh}(\omega + X) =&\, \omega^L_{\m\n\rh}(\omega) -\partial_{[\mu} \tr ( \omega_\n X_{\rh]} ) + \tr (R_{[\m\n} X_{\rh]}) 
\nn\w2
& +\tr (X_{[\m} D_\n(\omega) X_{\rh]} ) +\frac23 \tr ( X_{[\m} X_\n X_{\rh]} )\ .
\end{align}

%%%%%%%%%%%%%%%%%%%%%%%%%%%%%%%%%%%%%%%%%%%%%%%%%%%%%%%%%%%%%%%%%%
\section{Dimensional Reduction of $\cL_{BdR}^{10D, \rm dual}$ }
%%%%%%%%%%%%%%%%%%%%%%%%%%%%%%%%%%%%%%%%%%%%%%%%%%%%%%%%%%%%%%%%%%

Here we consider an ordinary dimensional reduction of the Lagrangian \eq{10DdualBS} on 4-torus, followed by a consistent truncation in which only kept bosonic fields are $(g_{\m\n}, B_{\m\n}, \vp)$, originating from
\be
{\hat g}_{\m\n} = g_{\m\n}\ ,\qquad {\hat\vp} = \vp\ ,\qquad  {\hat B}_{\m\n ijk\ell} = B_{\m\n} \e_{ijk\ell}\ ,
\ee
where the hatted fields are those of $10D$, and the unhatted are the ones in $6D$, and we have hatted the $10$ coordinates and split them as ${\hat x}^\mu= (x^\m, y^i),\ i=1,...,4$. Thus, recalling to set $\partial_i=0$, we get the intermediate results
\begin{align}
{\hat H}_{\m\n,\rh\s} &= H_{\m\n,\rh\s}\ ,
\qquad
{\hat H}_{\mu i,\nu j} = \tfrac12 \delta_{ij} H^2_{\m\n}\ ,
\qquad \hat{H}_{ij,kl}= \tfrac13 \delta_{i[k} \delta_{\ell]j} H^2\ ,
\nn\w2
{\hat H}^2_{\m\n} &= \tfrac12 H^2_{\m\n}\ ,\qquad {\hat H}^2_{ij} = \tfrac16 \delta_{ij} H^2\ ,
\qquad {\hat H}^2 = \tfrac16 H^2\ .
\label{nr1}
\end{align}
It follows that
\begin{align}
{\hat H}_4 &=  H_4 +2 H^{2\m\n} H^2_{\m\n} +\tfrac13 (H^2)^2\ ,
\nn\w2
{\hat H}^{2\hat\m\hat\n} {\hat H}^2_{\hat\m\hat\n} &= \tfrac14 H^{2\m\n} H^2_{\m\n} +\tfrac19 (H^2)^2\ .
\label{nr2}
\end{align}
Using these lemmas in the ordinary dimensional reduction of \eq{10DdualBS}, we get in $6D$
\begin{align}
\cL^{\rm dual}_{BdR} =&\, e\, e^{2 \vp} \Bigg[ \tfrac14 R(\omega) + g^{\m\n} \pd_\m \vp \pd_\n \vp - \tfrac{1}{12} e^{- 4 \vp} H_{\m\n\rh} H^{\m\n\rh} 
\nn\w2
&\, + \alpha'\,\Big( - \tfrac14 R^{\m\n\rh\s} R_{\m\n\rh\s} + 3 e^{- 4 \vp} R^{\m\n} H^2_{\m\n} - \tfrac12 e^{- 4 \vp} R H^2 
\nn\w2
&\, - \tfrac32 e^{- 4 \vp} R^{\m\n\rh\s} H_{\m\n, \rh\s}  - \tfrac16 e^{- 8 \vp} H_4 - \tfrac12 e^{- 8 \vp} H^2_{\m\n} H^{2 \m\n} 
\nn\w2
&\, + \tfrac13 e^{- 4 \vp} ( D_\m H_{\n\rh\s} ) D^\m H^{\n\rh\s}  + \tfrac43 e^{- 4 \vp} H^2 \vp^2 - \tfrac23 e^{- 4 \vp} \vp^\m D_\m H^2 
\nn\w2
&\, + e^{- 2 \vp} \tH^{\m\n\rh} \omega^L_{\m\n\rh}(\omega) +  e^{- 6 \vp}   \tH^{\m\n\rh} H_\m{}^{\al\beta} D_\n H_{\rh\al\beta} \Big) \Bigg] \ ,
\label{6DdualBS}
\end{align}
We have checked that this result agrees with direct dualization of $\cL_{BdR}$ in $6D$.

\end{appendix}

%%%%%%%%%%%%%%%%%%%%%%%%%%%%%%%%%%%%%%%%%%%%%%%%%%%%%%%%%%%%%%%%%%

\end{document}